\documentclass[twocolumn,pre,aps,showpacs,amsmath,amssymb,superscriptaddress]{revtex4}

\usepackage{hyperref}
\usepackage{amsmath,amssymb}
\usepackage{amsfonts,amsthm}
\usepackage{graphics}
\usepackage{graphicx}
\usepackage{dcolumn}
\usepackage{color}
\usepackage{bm}

\usepackage[normalem]{ulem}

\begin{document}

\title{A symbolic information approach to characterize
response-related differences in cortical activity during a Go/No-Go task}

\author{Helena B. Lucas}
\thanks{helena.bordini@fis.ufal.br}
\affiliation{Instituto de F\'{\i}sica, Universidade Federal de Alagoas, Macei\'{o}, Alagoas 57072-970 Brazil.}

\author{Steven L. Bressler}
\affiliation{Center for Complex Systems and Brain Sciences, Dept. of Psychology, Florida Atlantic University, Boca Raton, FL 33431, USA}

\author{Fernanda S. Matias}
\thanks{fernanda@fis.ufal.br}
\affiliation{Instituto de F\'{\i}sica, Universidade Federal de Alagoas, Macei\'{o}, Alagoas 57072-970 Brazil.}

\author{Osvaldo A. Rosso}
\thanks{oarosso@gmail.com}
\affiliation{Instituto de F\'{\i}sica, Universidade Federal de Alagoas, Macei\'{o}, Alagoas 57072-970 Brazil.}

\begin{abstract}
How the brain processes information from external stimuli in order to perceive the world and act on it is one of the greatest questions in neuroscience. To address this question different time series analyzes techniques have been employed to characterize the statistical properties of brain signals during cognitive tasks. Typically response-specific processes are addressed by comparing the time course of average event-related potentials in different trials type.
Here we analyze monkey Local Field Potentials data during visual pattern discrimination called Go/No-Go task in the light of information theory quantifiers. 
We show that the Bandt-Pompe symbolization methodology to calculate entropy and complexity of data is a useful tool to distinguish response-related differences between Go and No-Go trials. We propose to use an asymmetry index to statistically validate trial type differences.
Moreover, by using the multi-scale approach and embedding time delays to downsample the data we can estimate the important time scales in which the relevant information is been processed.
\end{abstract}
\maketitle

\section{Introduction}
\label{Sec-Introduction}
The brain is one of the biophysical systems that capture with greater attention our interest in understanding how it operates dynamically and organizationally~\cite{Sporns11}.
In a general description, one can think of the brain as an efficient complex network composed of many parts or structures and with an intense flow of information between these structures/zones. Moreover, this exchange of information could be considered non-linear and determine the corresponding brain behavior.
In particular, each one of these brain regions has its own structures and functions. They are also interconnected and share information between them, forming in this way an integrated network that determines the dynamics and behavioral response of the brain.

Taking these characteristics of the brain such as interconnections and information flow between different brain areas, measurement of time series activity at different regions and under external stimuli could bring us information about the corresponding dynamics, as well as about the information transfer. 
Integrated processes at the visual and motor systems during visual discrimination tasks have been extensively studied by analyzing monkey Local Field Potentials (LFP)  data~\cite{Bressler93,Liang00,Brovelli04,Salazar12,Dotson14}.
In particular, the averaged evoked response potential (ERP) of monkey LFP has been employed to estimate response-related differences between Go and No-Go trials~\cite{Ledberg07}.
The authors have shown that response-specific processing began around 
150~ms post-stimulus in widespread cortical areas.
The time course of task-related activity was examined to identify the cortical locations and the time windows that exhibit a significant difference between each response type. Here we propose that analysis of the statistical properties of the concatenated trials could also be useful and eventually could provide more information than comparing the activation response by averaging brain activity.

To do this type of study, we chose to work with time causal quantifiers based on Information Theory (Shannon entropy, MPR-statistical complexity, and entropy-complexity plane) \cite{Shannon49,Lamberti04,Rosso07,Zunino12,Chinitos20}.
These quantifiers are evaluated using the Bandt-Pompe symbolization methodology~\cite{Bandt02}, which includes naturally the time causal ordering provided by the time-series data in the corresponding associate probability distribution function (PDF). These tools have been employed to analyze brain signals in plenty of studies: to estimate time differences during phase synchronization~\cite{Montani15}, to show that complexity is maximized close to criticality in cortical states~\cite{Lotfi2020b},
to distinguish cortical states using EEG data~\cite{Rosso06} as well as
neuronal activity~\cite{Montani15b,Montani14}.

The manuscript is organized in the following way:
in Sec.~\ref{Sec:Information-quantifiers} we introduce the Information Theory quantifiers, as well as in Sec.~\ref{Sec:BP-method}, in which the Bandt-Pompe methodology for their evaluation is presented.
The experimental data corresponding to Local Field Potentials (LFP), are presented in Sec.~\ref{Sec:Experimental-data}, at four cortical-deep electrodes, in a monkey brain, and under visual stimulus condition Go/No-Go.
In Sec.~\ref{Sec:Results}, we report our analysis and results.
Finally, concluding remarks and a brief discussion of the significance of our findings for neuroscience are presented in Sec.~\ref{Sec:Conclusions}.

\section{Information-Theory quantifiers: permutation entropy and statistical complexity}
\label{Sec:Information-quantifiers}

In the characterization of dynamical systems (systems that evolve with time) the main objective is to infer some properties and behavior of the system under study. In particular, the starting point is a set of $M$ discrete measures taken at regular intervals of the same representative variable of the system, that we denote by the corresponding time series ${\mathcal X}(t)$.
In the second step, we need to associate a probability distribution function
$P$ to the time series, and with it, we will able to evaluate the corresponding Information Theory quantifiers. 

There is no unique answer for the best procedure to associate a time series with a PDF, and in fact, different proposals can be found. However, we start from a sequence/measurement taken in causality order, these characteristics are something to be preserved in the determination of the PDF.
For this reason, we follow the procedure proposed by Bandt-Pompe ~\cite{Bandt02} for the association of time causal PDF to the time series.

Let ${\mathcal X}(t)\equiv \{x_t; t=1,2,\dots,M\}$, be the time series representing a set of $M$ measurements of the observable ${\mathcal X}$ and let a probability distribution function be given by $P\equiv \{p_j;j=1,2,\dots,N\}$, with $\sum_{j=1}^{N} p_j=1$, and $0\leq p_j \leq 1$, where $N$ is the number of possible states of the system. 

The first Information Theory quantifier that we introduce is the Shannon's logarithmic information measure \cite{Shannon49}, defined by:
\begin{equation}
\label{eq:Shannon-entropy}
S[P] = -\sum_{j=1}^{N} p_j \ln(p_j) \ .
\end{equation}
This functional is equal to zero when we are able to correctly predict the outcome every time. For example, for linearly increasing time series all probabilities are zero but one which is equal to 1. The corresponding PDF will be $P_0 = \{ p_k =1 ~{\mathrm {and}} ~p_j = 0, \forall j\neq k, j = 1, \dots, N-1 \}$, then $S[P_0]=0$. 
By contrast, the entropy is maximized for the uniform distribution $P_e = \{p_j=1/N, \forall j=1,2, \dots,N\}$, being $S_{max}= S[P_e]= \ln( N )$.
We define the normalized Shannon entropy by
\begin{equation}
\label{eq:Shannon-entropy}
H[P] = \frac{S[P]}{S_{max}} = \frac{S[P]}{\ln(N)} \ ,
\end{equation}
and $0 \leq H[P]\leq 1$, which give a measure of the information content of the corresponding PDF ($P$).

The second Information Theory based quantifier that we introduced is the Statistical Complexity, defined by functional product form ~\cite{LMC95}
\begin{equation}
\label{eq:Complexity}
C[P, P_e] = H[P] \cdot  Q_J[P,P_e] \ .
\end{equation}
Where, $H[P]$ is the normalized Shannon entropy (see Eq.~(\ref{eq:Shannon-entropy})) and $Q_J[P,P_e]$ represent the disequilibrium, which is defined in terms of the Jensen–Shannon divergence ~\cite{Grosse02} as:
\begin{equation}
\label{eq:Q}
Q_J[P,P_e] = Q_0 J[P,P_e] \ ,
\end{equation}
where
\begin{equation}
\label{eq:C}
J[P,P_e] = S\left[\frac{(P+P_e)}{2}\right] - \frac{S[P]}{2} - \frac{S[P_e]}{2}  \ ,
\end{equation}
and $Q_0$ is a normalization constant ($0\leq Q_J \leq 1$), equal to the inverse of the maximum possible value of $J[P,P_e]$, that is $Q_0 = 1 / J[P_0,P_e]$.
In this way, also the statistical complexity is a normalized quantity, $0\leq C[P,P_e] \leq 1$.
It is interesting to note, that the statistical complexity give additional information in relation to the entropy, due to its dependence on two PDF, moreover, can be shown that for a given value of the normalized entropy $H$, the corresponding complexity varies in a range of values given by $C_{\mathrm{min}}$ and $C_{\mathrm{max}}$, and these values depend only on the dimension of the PDF considered and the functional form chosen for the entropy ~\cite{MPR-cotas}.

\begin{figure}[t]
\includegraphics[width=0.95\columnwidth,clip]{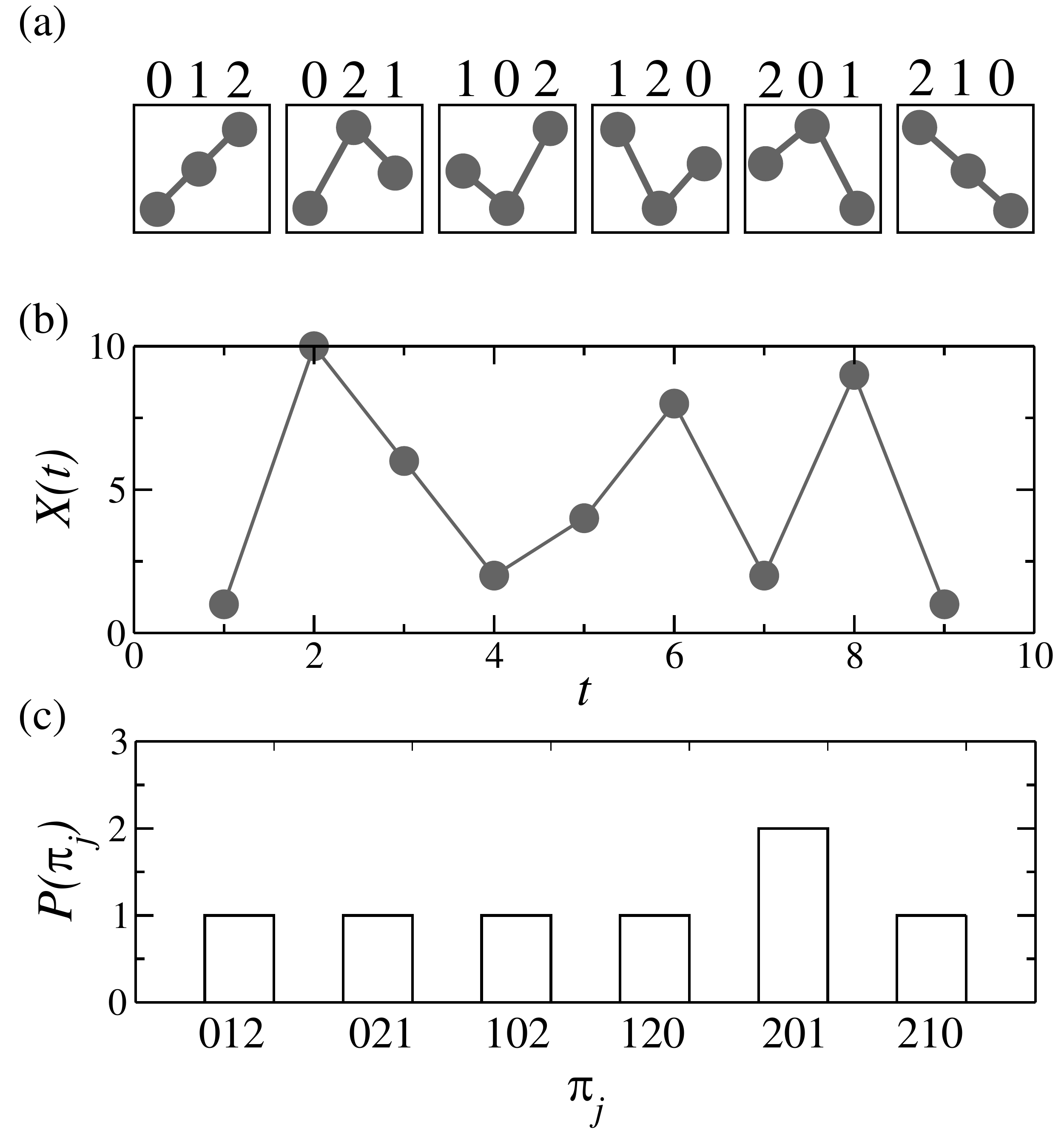}
\caption{  
Characterizing the symbolic information approach.
(a) The six symbols associated to permutations $\pi_j$ for ordinal patterns of length $D=3$, and embedding time $\tau=1$.
(b) Simple example of a time series given by $X(t)=\{1,10,6,2,4,8,2,9,1\}$, $M=9$ and, 
(c) its own non-normalized probability density function (PDF).
}
\label{fig:symbol}
\end{figure}

\section{The Bandt-Pompe symbolization method}
\label{Sec:BP-method}

Let ${\mathcal X}(t) \equiv \{ x_t; t=1, \dots, M \}$ denote a time series measured from the dynamics state of a dynamical system. $M$ is the number of data (time series length) measured at regular equal spaced times.
We define the symbol vector ${\vec{\pi}}^{(D)}$ by
\begin{equation}
{\vec{\pi}}^{(D)}~=~ (\pi_0, \pi_1, \pi_2, ~\dots~, \pi_{D-1} ) \ ,
\label{eq:pi-vector}
\end{equation}
in such a way that every element of ${\vec{\pi}}^{(D)}$ is unique 
($\pi_j \neq \pi_k$ for every $j \neq k$).

Given the parameters $D \in \mathbb{N}$ with $D \geq 2$, the embedding dimension which represent the quantity of information to be included and; the parameter $\tau \in \mathbb{N}$ with $\tau \geq 1$, the embedding time (time delay),  we divide the time series ${\mathcal X}(t)$ in 
overlapping vectors of length $D$,
\begin{equation}
{\vec{Y}}_{s}^{(D,\tau)}~=~ (x_s, x_{s+\tau}, x_{s+2\tau}, ~\dots~, 
                                  x_{s+(D-1)\tau} )  \ ,
\label{eq:Y-vector}
\end{equation}
with $s=1, 2, \dots, M-(D-1)\tau$.
The index $s$ controls the beginning of each vector and $\tau$ the overlap degree between vectors.
The vector ${\vec{Y}}_{s}^{(D,\tau)}$ can be mapped to a symbol vector ${\vec{\pi}}^{(D)}_s$. 
This mapping is such that preserves the desired relation between the elements $x_s \in {\vec{Y}}_{s}^{(D,\tau)}$ and all $s \in \{1, 2, \dots, M-(D-1) \tau \}$ that share this pattern (also called motif) and are mapped to the same ${\vec{\pi}}^{(D)}$.
We define the mapping overlapping vectors of length $D$,
\begin{equation}
{\vec{Y}}_{s}^{(D,\tau)} ~\mapsto~ {\vec{\pi}}_{s}^{(D)}   \ ,
\label{eq:maping}
\end{equation}
by ordering the observations $x_s \in {\vec{Y}}_{s}^{(D,\tau)}$ in increasing order.
That is, from the vector ${\vec{Y}}_{s}^{(D,\tau)}$ (see eq.~(\ref{eq:Y-vector}) 
we looking for the vector
\begin{equation}
{\vec{\pi}}_{s}^{(D)} ~=~(r_0, r_1, r_2, ~\dots~, r_{(D-1)}) \ ,
\label{eq:pi-vector}
\end{equation}
which represent the permutation pattern, given by
\begin{equation}
{\pi}_{s}^{(D)} ~=~\left[0, 1, 2, ~\dots~, (D-1) \right] \ ,
\label{eq:pi-pattern}
\end{equation}
such that
\begin{equation}
x_{s+r_0} < x_{s+r_1} < x_{s+r_2} < ~\dots~ < x_{s+r_{D-1)}} \ .
\label{eq:order-componentes}
\end{equation}
In order to get an unique result, we set $r_j < r_{j+1}$ if
$x_{s+r_j} = x_{s+r_{j+1}}$.

The Bandt-Pompe probability distribution function (BP-PDF) of ordinal patterns $\Pi^{D}$ is obtained by the frequency histogram of symbols associated to a given time series ${\vec{Y}}_{s}^{(D,\tau)}$.
The BP-PDF is the relative frequency of symbols in the series against the $D!$ possible patterns, that is
 \begin{equation}
p(\pi_s^{(D)})~=~\frac{\sharp\{ {\vec{Y}}_{s}^{(D,\tau)} 
                            {\textrm{ is of type }} 
                            {{\pi}}_{s}^{(D)} \}
                   }{[M-(D-1)\tau]} ~    \ ,
\label{eq:maping}
\end{equation}
in which the symbol $\sharp$ indicate number, and 
$s=\{1,2, \dots, M-(D-1)\tau \}$. 
Note that $p(\pi_s^{(D)})$ satisfies the probability condition 
$0\leq p(\pi_s^{(D)}) \leq 1$ and $\sum_{j=1}^{D!} p(\pi_s^{(D)}) =1$.
An important property of the BP-PDF is its invariant monotonic transformations.

Let consider an example: given the time series
${\mathcal X}(t)=\{1,10,6,2,4,8,2,9,1\}$, with $M=9$, we evaluate the BP-PDF with $D=3$ and $\tau =1$. 
In Fig.~\ref{fig:symbol}.a, we represent the 6 ordinal patterns, corresponding to $D=3$,
and in the Fig.~\ref{fig:symbol}.b, the time series considered is represented as function of time.
Then we have $N' = M - (D-1) \tau = 7$ embedding vectors: \\
${\vec{Y}}^{(3,1)}_1=(1,10,6)~\mapsto~\vec{\pi}^{(3)}_1=(0,2,1) 
\mapsto \pi_2 = [021]$;\\
${\vec{Y}}^{(3,1)}_2=(10,6,2)~\mapsto~\vec{\pi}^{(3)}_2=(2,1,0) 
\mapsto \pi_6 = [210]$; \\
${\vec{Y}}^{(3,1)}_3=(6,2,4)~\mapsto~\vec{\pi}^{(3)}_3=(1,2,0) 
\mapsto \pi_4 = [120]$; \\
${\vec{Y}}^{(3,1)}_4=(2,4,8)~\mapsto~\vec{\pi}^{(3)}_4=(0,1,2) 
\mapsto \pi_1 = [012]$; \\
${\vec{Y}}^{(3,1)}_5=(4,8,2)~\mapsto~\vec{\pi}^{(3)}_5=(2,0,1) 
\mapsto \pi_5 = [201]$; \\
${\vec{Y}}^{(3,1)}_6=(8,2,9)~\mapsto~\vec{\pi}^{(3)}_6=(1,0,2) 
\mapsto \pi_3 = [102]$; \\
${\vec{Y}}^{(3,1)}_7=(2,9,1)~\mapsto~\vec{\pi}^{(3)}_7=(2,0,1) 
\mapsto \pi_5 = [201]$. \\
Then the pattern probability are $p(\pi_1)=p(\pi_2)=p(\pi_3)=p(\pi_4)=p(\pi_6)=1/7$ and $p(\pi_5)=2/7$.
Fig.~\ref{fig:symbol}.c shows the non-normalized histogram patterns.

\begin{figure}[t]%
 \begin{minipage}{8cm}
  \begin{flushleft}(a)%
\end{flushleft}%
\includegraphics[width=0.98\columnwidth,clip]{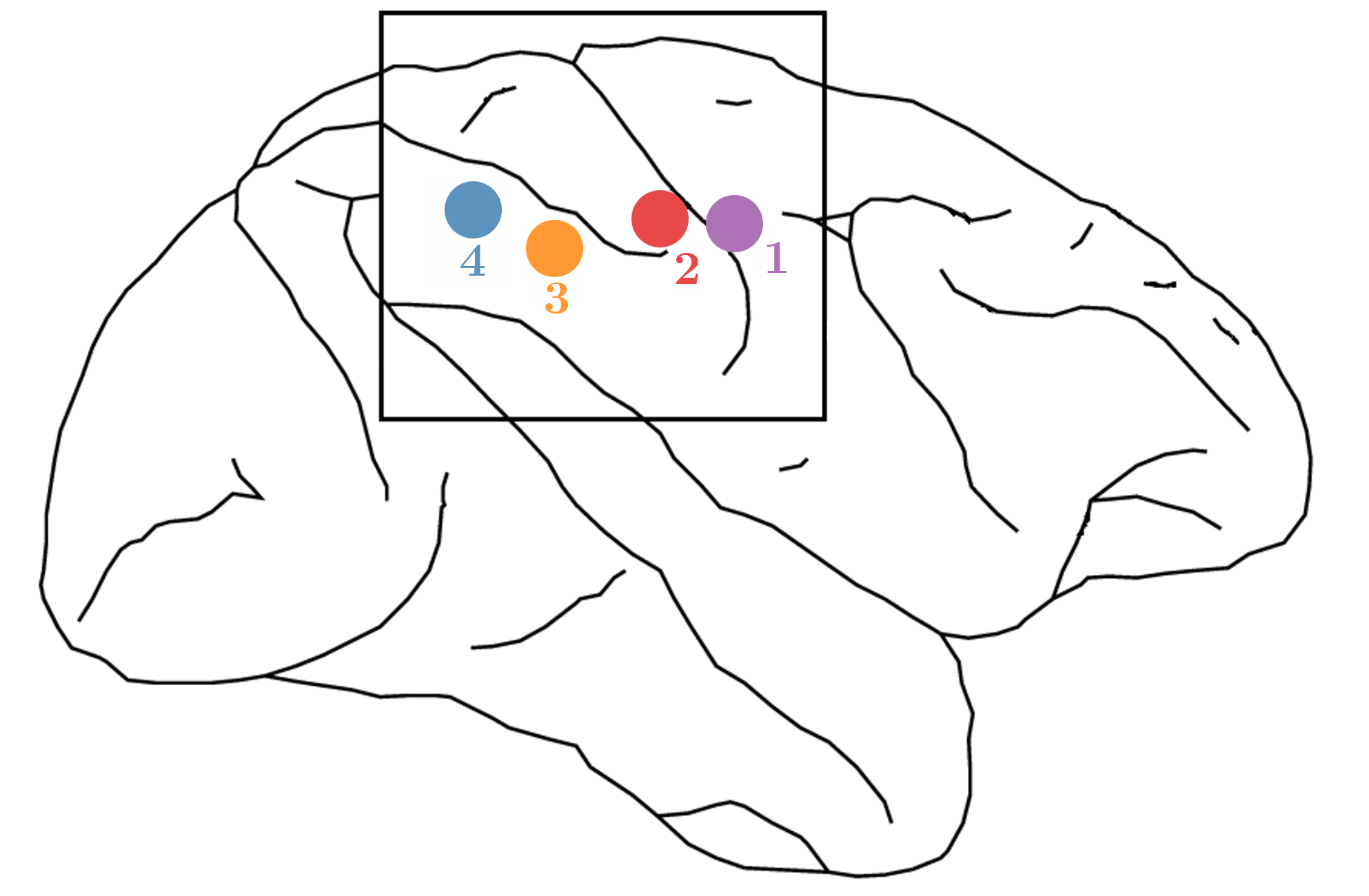}
\end{minipage}
\begin{minipage}{8cm}
\begin{flushleft}(b)%
\end{flushleft}%
\includegraphics[width=0.98\columnwidth,clip]{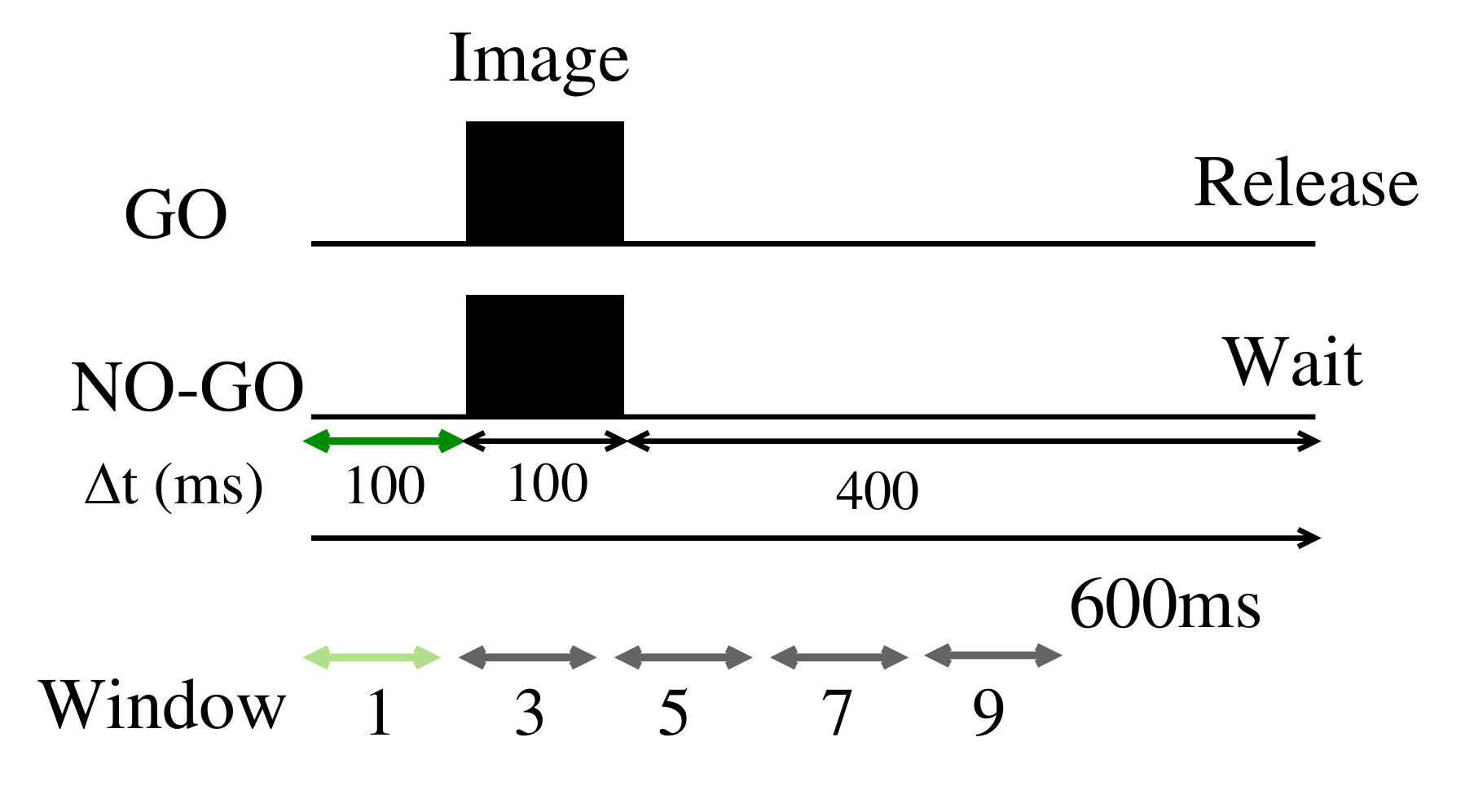} 
\end{minipage}
\caption{\label{fig:motif} 
Schematic representation of the experimental setup. 
(a) Electrodes position in the monkey brain: primary motor cortex (channel 1, violet), primary somatosensory  cortex (channel 2, red), posterior parietal areas (site 3 in orange and 4 in blue). (b) Temporal organization of the Go/No-Go task. We analyze the statistical properties of the time series in different intervals of the trial by separating it in 11 windows ($W_i$) of $90$~ms each, starting every $50$~ms. Each trial goes from $t=0$ to $t=600$~ms (see more details in Sec.~\ref{Sec:Experimental-data}).
  }
\end{figure}%

\section{Experimental time series}
\label{Sec:Experimental-data}

\begin{figure}[t]%
 \includegraphics[width=0.9\linewidth,clip]{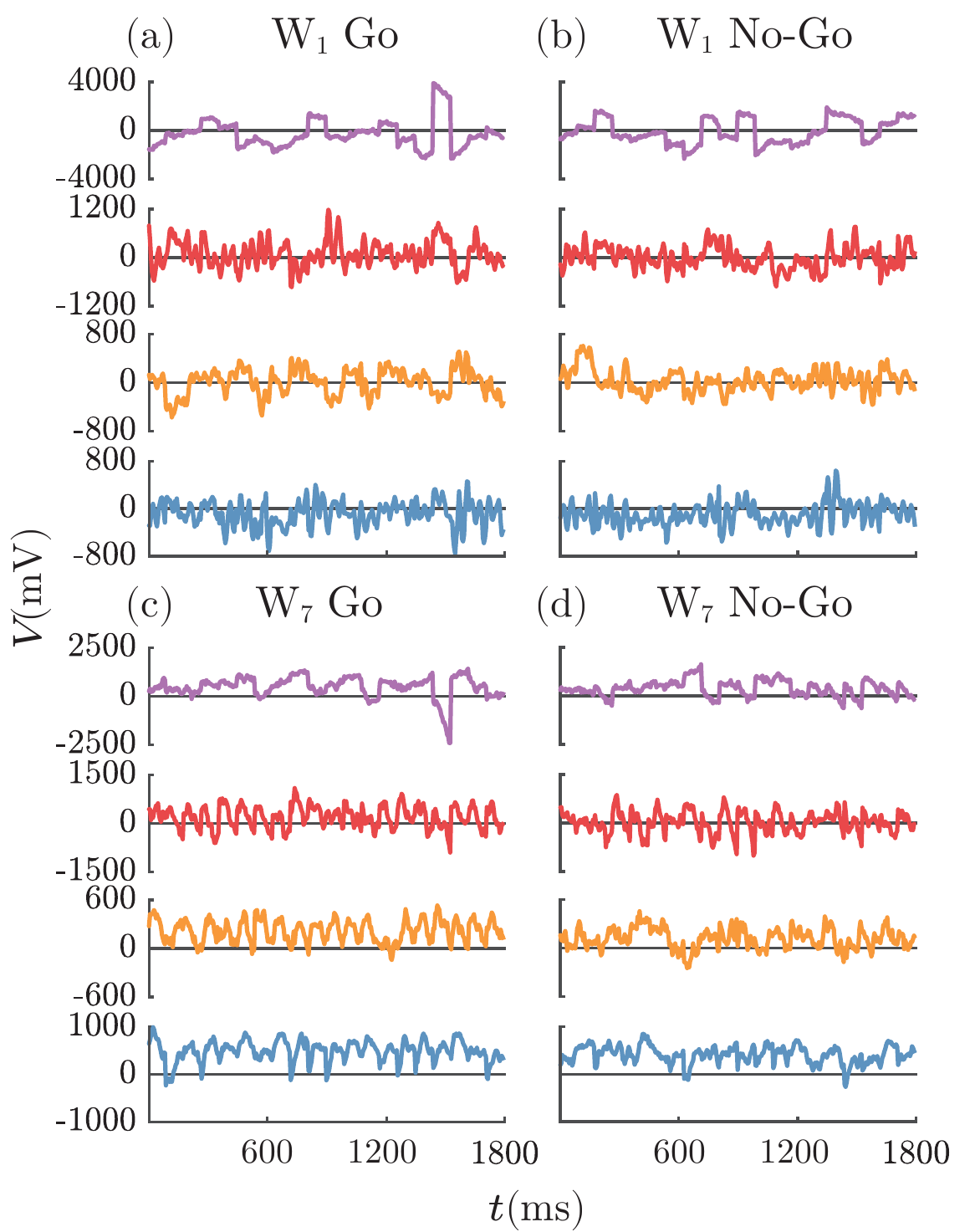}
\caption{\label{fig:timeseries} 
Illustrative examples of the time series from monkey-LFP taken over concatenated first 20 trials for the analyzed regions primary motor cortex;  somatosensory cortex; parietal cortices. Same color code for the signals from each region as depicted in Fig~\ref{fig:motif}(a). Time window $W_1$ ($0<t<90)$~ms) (a) under Go condition and (b) under No-Go condition. Time window $W_7$ ($300<t<390$~ms) (c) Go trials and (d) for No-Go trials.
}
\end{figure}

Local Field Potentials were recorded surface-to-depth with bipolar Teflon-coated platinum micro-electrodes from 15 distributed sites located in the right hemisphere of the adult male rhesus macaque monkeys (GE). 
Here we analyze the four sites shown in Fig.~\ref{fig:motif}(a). 
Site 1 is in the primary motor cortex, site 2 in the primary somatosensory cortex, and sites 3 and 4 are in the posterior parietal cortex. 
LFPs time series were sampled at 200 points/s (every $5$~ms), and collected from $100$~ms before to $500$~ms after stimulus onset (see Fig.~\ref{fig:timeseries} for illustrative examples of LFPs). Therefore we define our trial time from $t=0$ to $t=600$~ms, which means that the stimulus appears at $t=100$~ms.

The monkey was highly trained to perform a visual pattern discrimination task called Go/No-Go.  
On each trial, the monkey depressed a hand lever and kept it pressed during
a random interval ranging from $0.12$ to $2.2$~s while waiting for stimulus appearance. 
On Go trials, a water reward was provided if the monkey released the lever within $500$~ms after stimulus onset (time trial $t=600$~ms). On No-Go trials the monkey should not release the lever.
Experiments were performed at the Laboratory of Neuropsychology at the National Institute of Mental Health (USA), and animal care was in accordance with institutional guidelines at the time.

We separate each trial in 11 windows ($W_i$) of $90$~ms each, starting every $50$~ms. For example, the first window $W_1$ ranges from $0$ to $90$~ms, $W_2$ is from $50$ to $140$~ms, $W_3$ is from $100$ to $190$~ms (as illustrated in Fig.~\ref{fig:motif}(b) for the even windows). 
The average response for Go trials (mean
and standard deviation over sessions) occurs at
$t=349\pm9$~ms~\cite{Ledberg07} (which corresponds to a response time of $249\pm9$ ms since our zero time is set $100$~ms before the stimulus onset).
We concatenate the points from each $W_i$ window of all 359 Go trials and, separately, we concatenate all 351 No-Go trials. Figures~\ref{fig:timeseries}(a) and (b) show the first 20 Go trials (whereas figures ~\ref{fig:timeseries}(c) and (d) show the first 20 No-Go trials) of each region of interest during $W_1$ and $W_7$. We use this concatenated time series to extract the symbols and calculate the PDF in order to obtain the entropy and the complexity.

The first window has been previously analyzed in the light of the entropy-complexity plane by Montani et al.~\cite{Montani15} and by using Granger causality measures~\cite{Brovelli04,Matias14} to infer the connectivity between these regions. The entire task has been analyzed using coherence and Granger causality and separating Go and No-Go trials for sites 1, 2, and 3 by Zhang et al.~\cite{Zhang08}. The same sites and a few others have been analyzed in the light of Evoked Response Potentials (ERP) to study stimulus-evoked activation onset, stimulus-specific processing, stimulus category-specific processing, and response-specific processing by Ledberg et al.~\cite{Ledberg07}.

\begin{figure}%
 \includegraphics[width=0.9\linewidth,clip]{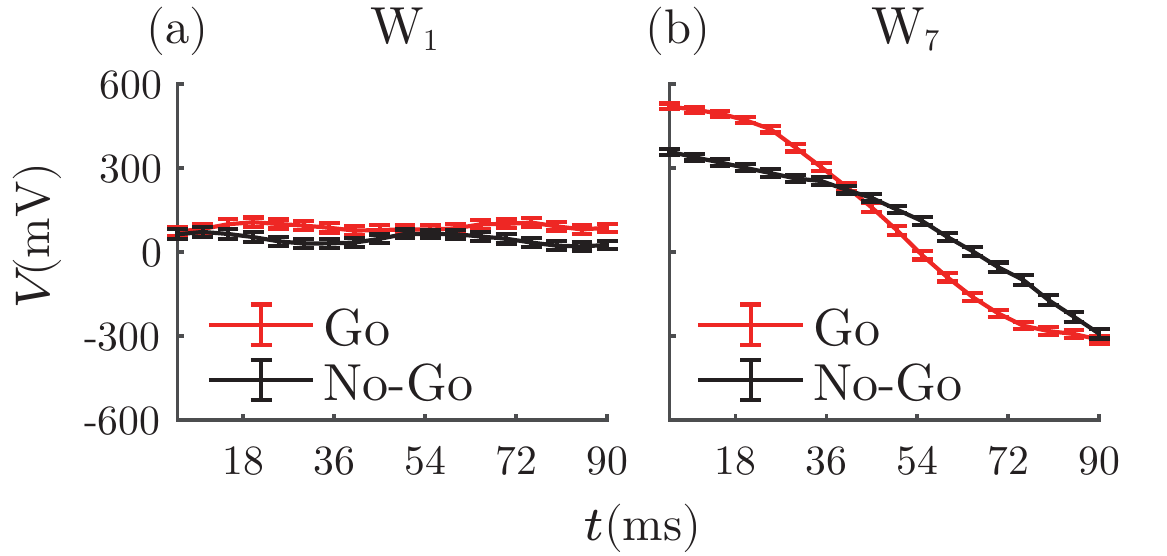}
\caption{\label{fig:mediaseries}
Comparison between Go and No-Go trials. Time course of average event-related potentials (ERP) in region 2: (a) over all corresponding trials for time window $W_1$ (pre-stimulus)  and (b) over all corresponding trials for time window $W_7$ (post-stimulus). Error bars are the standard deviation of the mean.
}
\end{figure}%

\section{Results}
\label{Sec:Results}

Here we employ information theory quantifiers as a useful tool to study response-specific processing in brain signals during a visual-motor task.
We have separately analyzed the monkey-LFP time series of all Go response trials, as well as, of all No-Go response trials for the four brain regions shown in Fig.~\ref{fig:motif}(a): (1) the primary motor cortex, (2) the somatosensory cortex, (3) and (4) the parietal cortex (see more details in Sec.~\ref{Sec:Experimental-data}). Illustrative examples of the time series for each one of the four sites during $W_1$ (from trial time $t=0$ to $90$~ms) and $W_7$ (from trial time $t=300$ to $390$~ms) for the first 20 trials of Go and No-Go conditions are shown in Fig.~\ref{fig:timeseries}.  
The activity of the motor cortex is clearly different from the other three regions. Moreover, in a naive comparison, just by eye inspection, it seems that Go and No-Go trials are more similar during $W_1$ than during $W_7$ in the four channels.

\begin{figure}[!t]%
 \begin{minipage}{8cm}
 \includegraphics[width=0.9\columnwidth,clip]{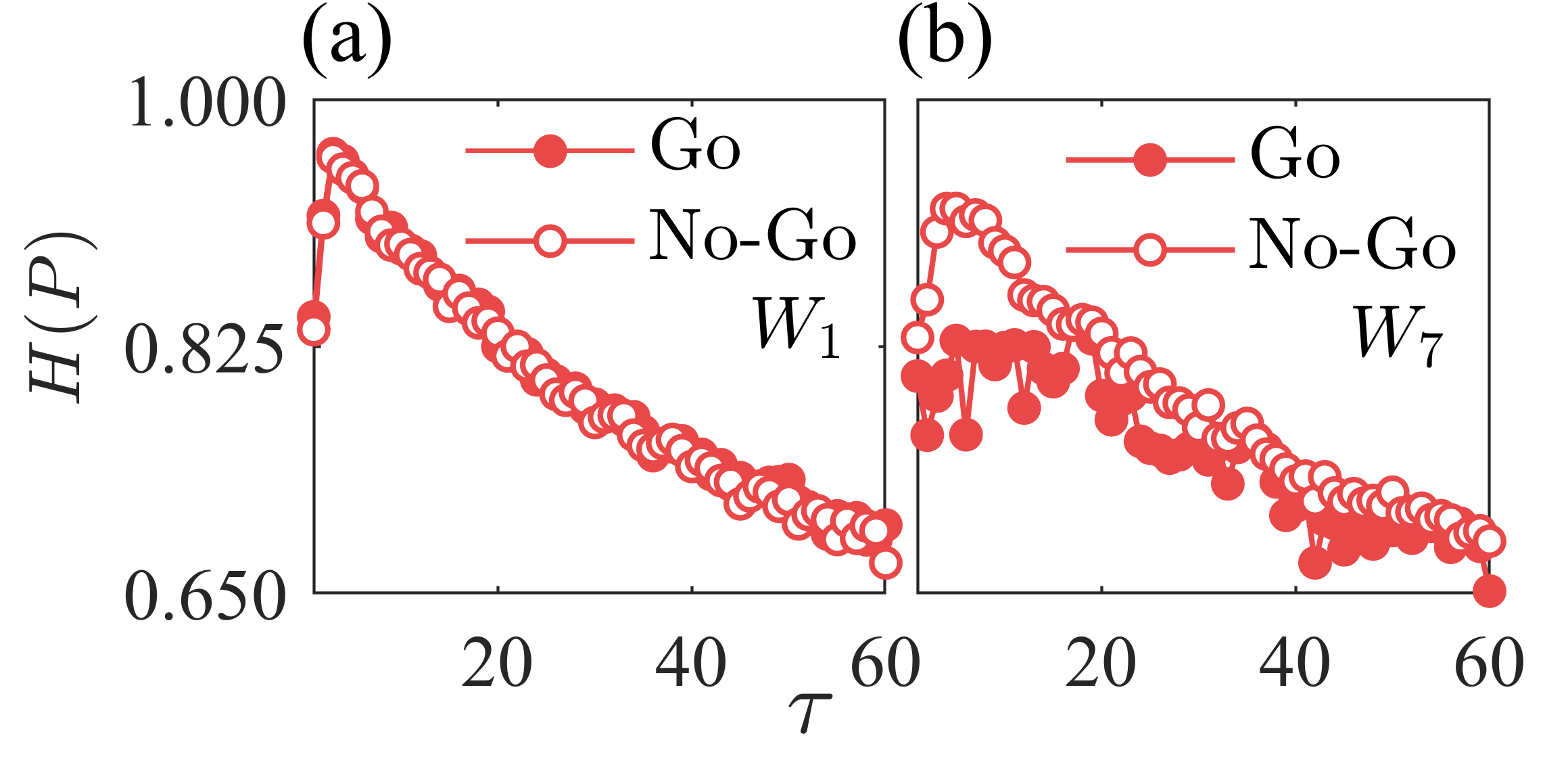}
\end{minipage}
 \begin{minipage}{8cm}
 \includegraphics[width=0.9\columnwidth,clip]{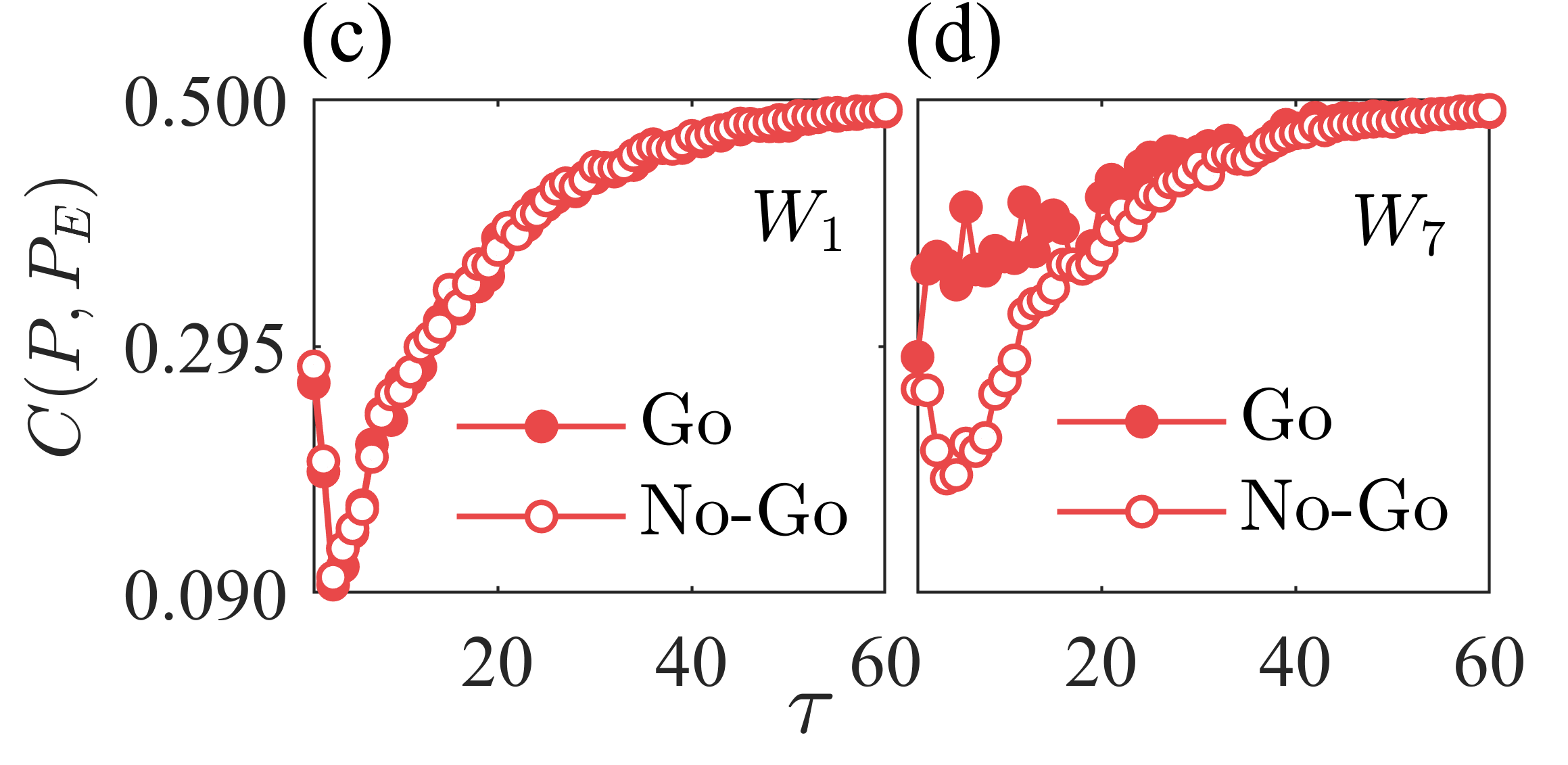}
\end{minipage}
\caption{\label{fig:CxTau_reg2}  Response-specific differences in region 2 captured by the Information theory quantifiers. Entropy as a function of the embedding time (or time delay) $\tau$ for (a) $W_1$ and (b) $W_7$. Complexity as a function of the embedding time $\tau$ for (c) $W_1$ and (d) $W_7$. 
}
\end{figure}%

Our first hypothesis is related to the fact that for the first two time windows ($W_1$ and $W_2$) 
 one should not find any statistically significant difference between the Go and No-Go time series in any region since the visual stimulus did not appear or has not been processed yet. However, before the end of the trial, we should be able to distinguish both cases. 
For example, in Fig.~\ref{fig:mediaseries}(a) and (b) we compare the average activity of all Go and all No-Go trials of the primary somatosensory area (region 2). The difference between the types of trials is clearly larger for $W_7$ than for $W_1$. This approach has been employed by Ledberg et al.~\cite{Ledberg07} along the whole trial to estimate how long it takes for each region to start to show significant differences between Go and No-Go conditions. In fact, Fig.~\ref{fig:mediaseries}(b) is comparable to Fig. 12(G) in Ref.~\cite{Ledberg07}. They have shown that this area starts to process response-specific information after $t=300$~ms which coincides with the beginning of $W_7$.
In what follows, we show that response-related differences can also be verified with the information theory quantifiers.

To characterize different cortical states during the cognitive task we have calculated the entropy and complexity for each $W_i$ in different time scales by changing $\tau$ from $1$ up to $60$, which corresponds to downsample the series from every $5$~ms up to every $300$~ms.
In Fig.~\ref{fig:CxTau_reg2} we show $H$ \textit{versus} $\tau$ and $C$ \textit{versus} $\tau$ for region 2 during $W_1$ and $W_7$. As expected, there is no difference between Go/No-Go trials for $W_1$. However, for $W_7$ both the entropy and the complexity are clearly different when comparing Go and No-Go trials. In particular, both conditions have an increase in the minimal complexity and a decrease in the maximal entropy when compared with the first window, but changes in the Go condition are much more pronounced. This is a good indication that we can use $H$ and $C$ together with the average potential to infer if the region is processing response-specific information. In all regions, differences between response types, when they exist, are much more pronounced for $\tau$ up to 15 ($75$~ms).

\begin{figure}
 \begin{minipage}{4cm}
  \begin{flushleft}(a)%
\end{flushleft}%
\includegraphics[width=0.98\columnwidth,clip]{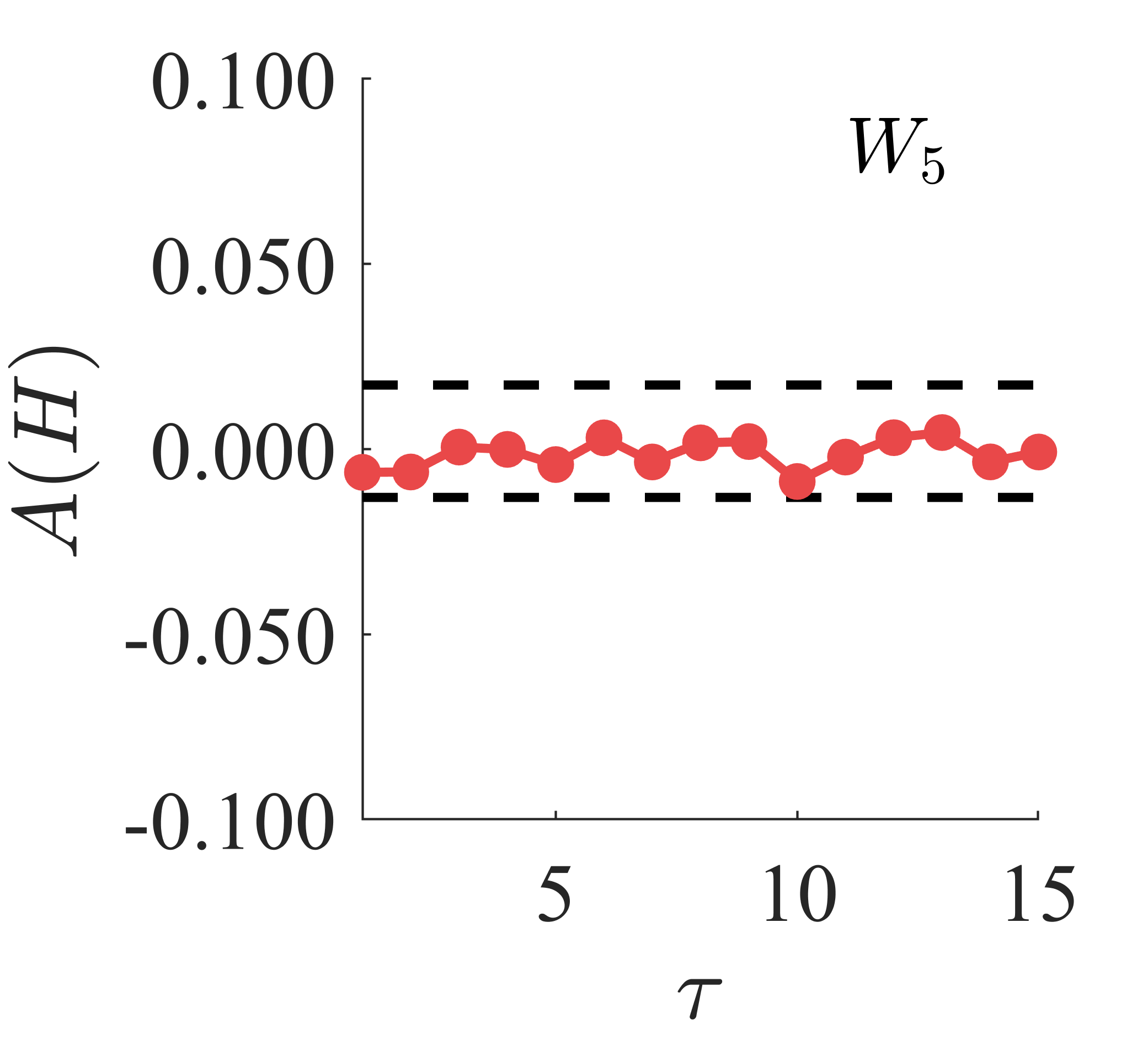}
\end{minipage}
\begin{minipage}{4cm}
\begin{flushleft}(b)%
\end{flushleft}%
\includegraphics[width=0.98\columnwidth,clip]{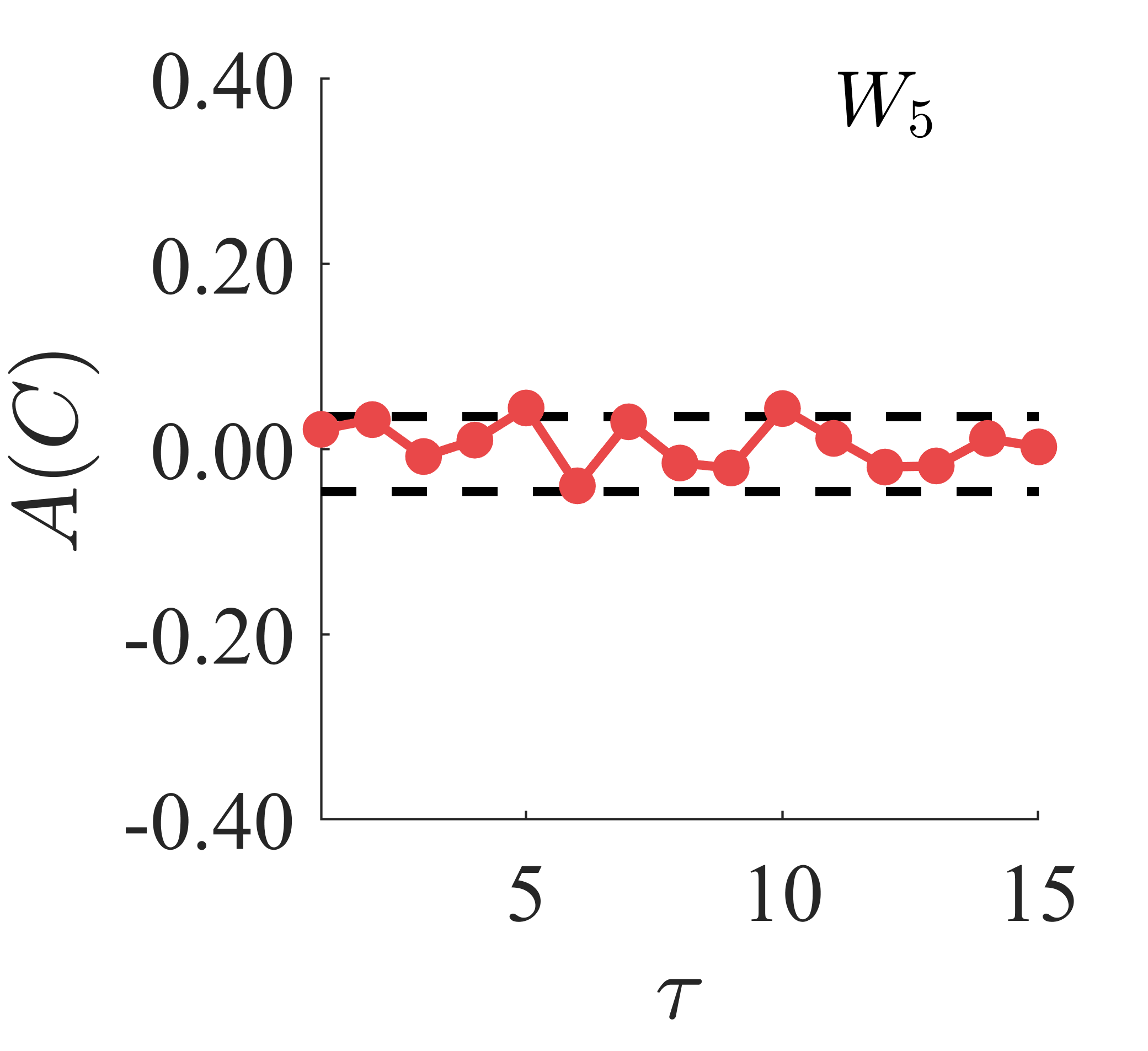}
\end{minipage}
 \begin{minipage}{4cm}
  \begin{flushleft}(c)%
\end{flushleft}%
\includegraphics[width=0.98\columnwidth,clip]{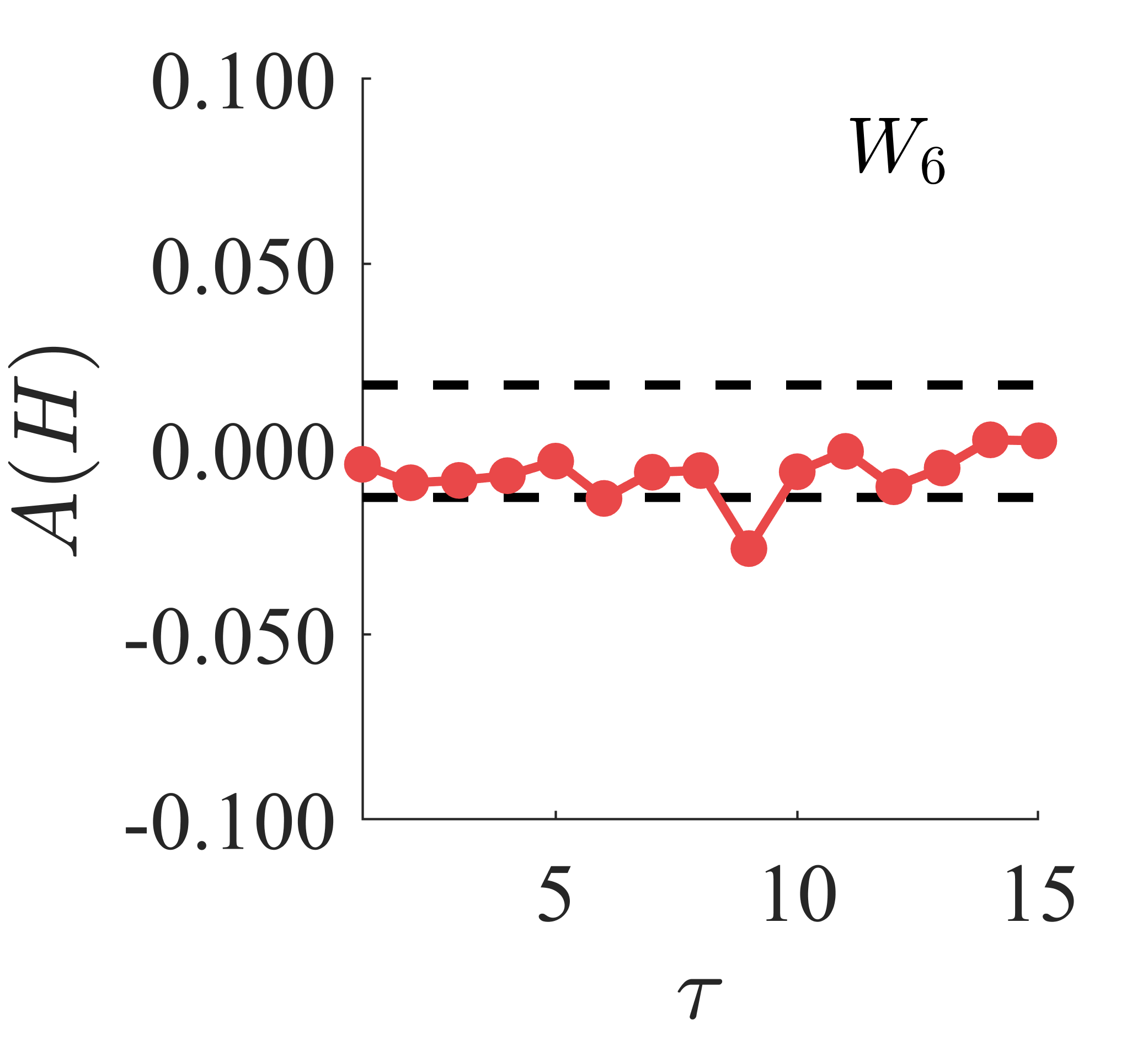}
\end{minipage}
\begin{minipage}{4cm}
\begin{flushleft}(d)%
\end{flushleft}%
\includegraphics[width=0.98\columnwidth,clip]{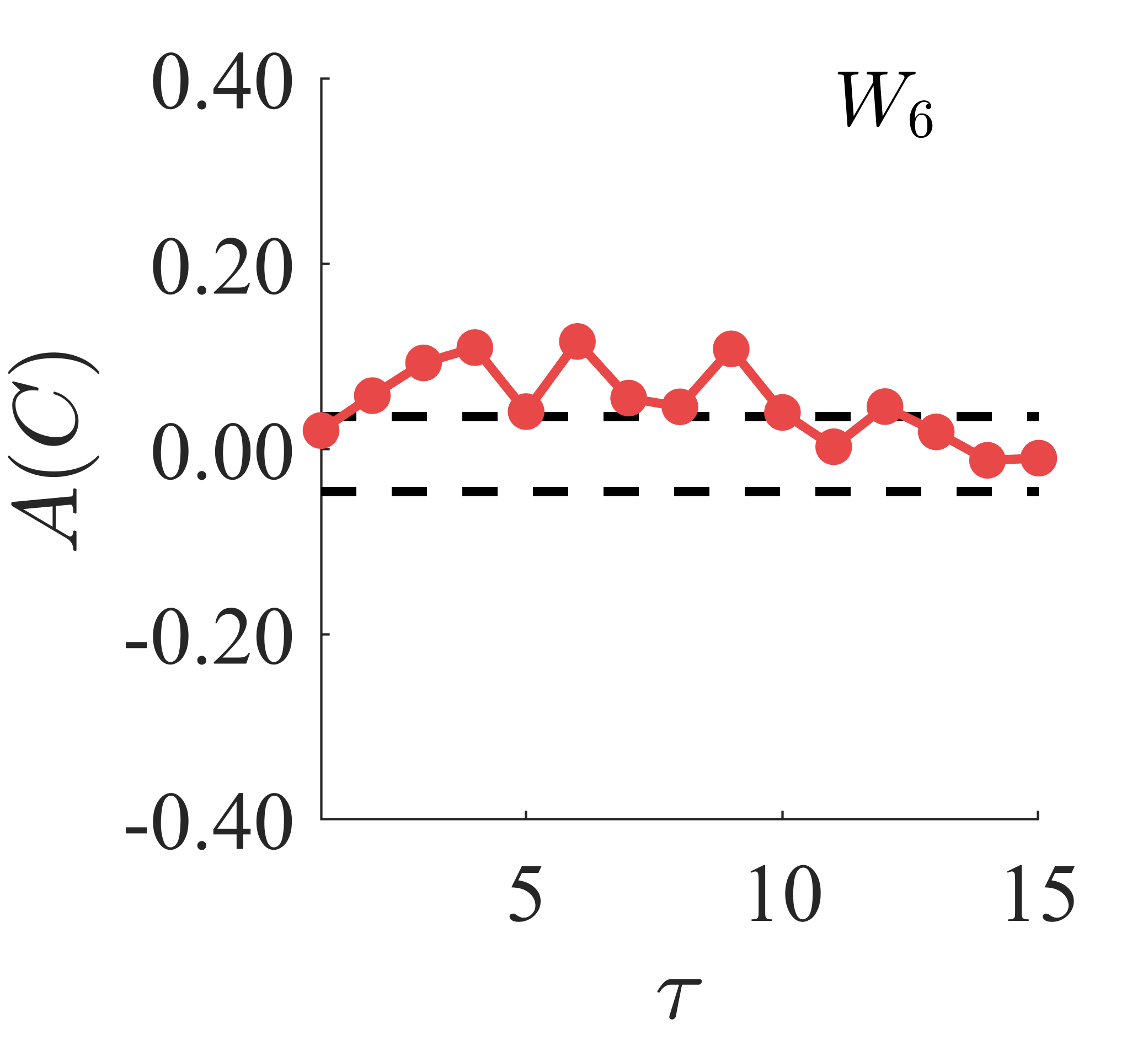}
\end{minipage}
 \begin{minipage}{4cm}
  \begin{flushleft}(e)%
\end{flushleft}%
\includegraphics[width=0.98\columnwidth,clip]{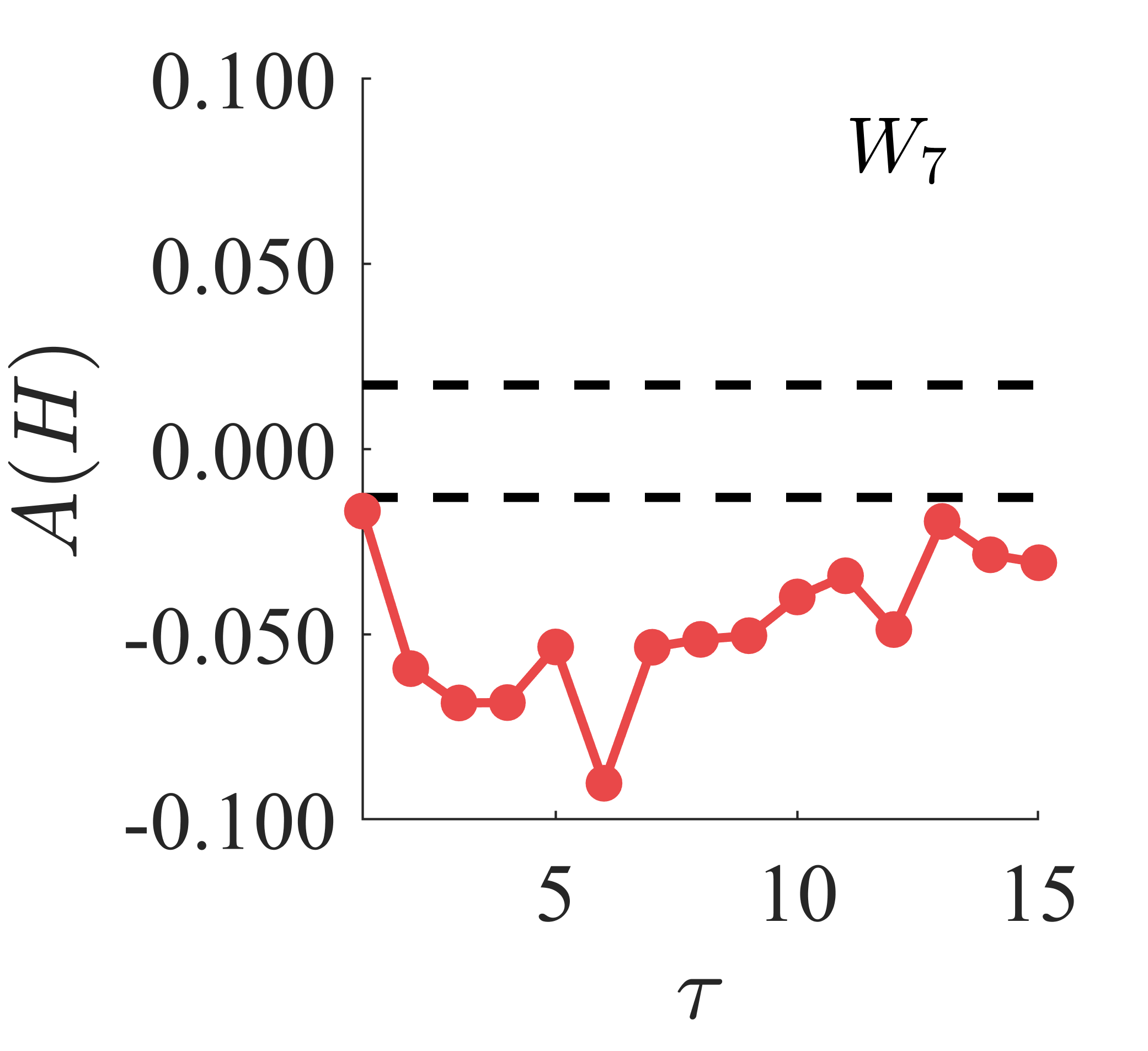}
\end{minipage}
\begin{minipage}{4cm}
\begin{flushleft}(f)%
\end{flushleft}%
\includegraphics[width=0.98\columnwidth,clip]{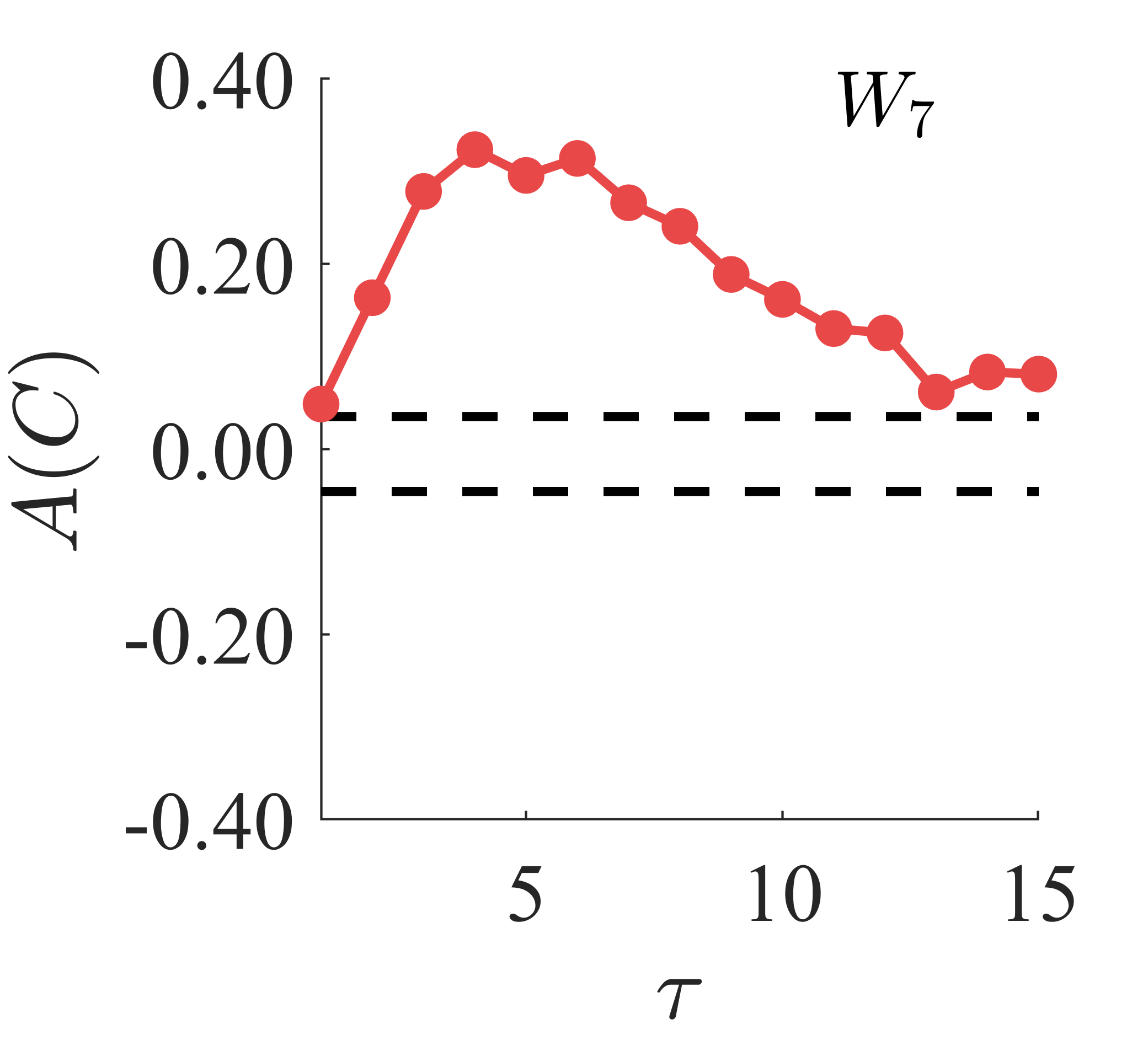}
\end{minipage}
\caption{\label{fig:AH_AC_CH_reg2}
Quantifying the statistical differences between Go and No-Go trials.
Asymmetry index for entropy $A(H)$ (left column) and complexity $A(C)$ (right column) for region 2 as a function of the time delay $\tau$ for three consecutive time windows (a, b) $W_5$, (c, d) $W_6$ and (e, f) $W_7$. We use $W_1$ and $W_2$ to calculate the average value $\mu_{A(C)}$ ($\mu_{A(H)}$) and its standard deviation $ \sigma_{A(C)}$ ($ \sigma_{A(H)}$). Dashed lines represent $A(H) = \mu_{A(H)} \pm 3 \sigma_{A(H)}$ and $A(C) = \mu_{A(C)} \pm 3 \sigma_{A(C)}$.
}
\end{figure}%

In order to quantify the response-specific difference, and statistically validate it,
we define an asymmetry index for entropy $A(H)$ and complexity $A(C)$ respectively by:
\begin{equation}
\label{asymmetric-H}
    A(H)=\frac{ H_{Go} - H_{NoGo}} {H_{Go} + H_{NoGo}} 
\end{equation}
and
\begin{equation}
\label{asymmetric-C}
    A(C)=\frac{ C_{Go} - C_{NoGo}} {C_{Go} + C_{NoGo}}  \ .
\end{equation}
We calculate $A(H)$ and $A(C)$ for each $\tau$, region and window. For each region we use $W_1$ and $W_2$ to calculate the average value  $\mu_{A(C)}$ ($\mu_{A(H)}$) and its standard deviation $ \sigma_{A(C)}$ ($ \sigma_{A(H)}$). We consider that the difference in the entropy (complexity) between Go and No-Go trials is significant if $A(H)> \mu_{A(H)} \pm 3 \sigma_{A(H)}$ ($A(C)> \mu_{A(C)} \pm 3 \sigma_{A(C)}$), see dashed lines in Fig.~\ref{fig:AH_AC_CH_reg2}.

In Fig.~\ref{fig:AH_AC_CH_reg2} we show these asymmetries as a function of the time delay in region 2 for three consecutive time windows: $W_5$, $W_6$ and $W_7$.
From $W_1$ to $W_5$ the results are very similar and it is not possible to verify significant differences between trial types with neither entropy nor complexity during these windows. However, at $W_6$ (from trial time $t=250$ to $340$~ms) the response-specific differences in complexity start to appear for intermediate values of time delay. This also corroborates the importance of using not only the entropy. At $W_7$ the asymmetry can be verified for both $H$ and $C$ at all time scales.
In particular, the Go/No-Go difference at the somatosensory region is larger for $W_7$ than for all others $W_i$.

\begin{figure}[t]%
 \begin{minipage}{4cm}
  \begin{flushleft}(a)%
\end{flushleft}%
\includegraphics[width=0.99\columnwidth,clip]{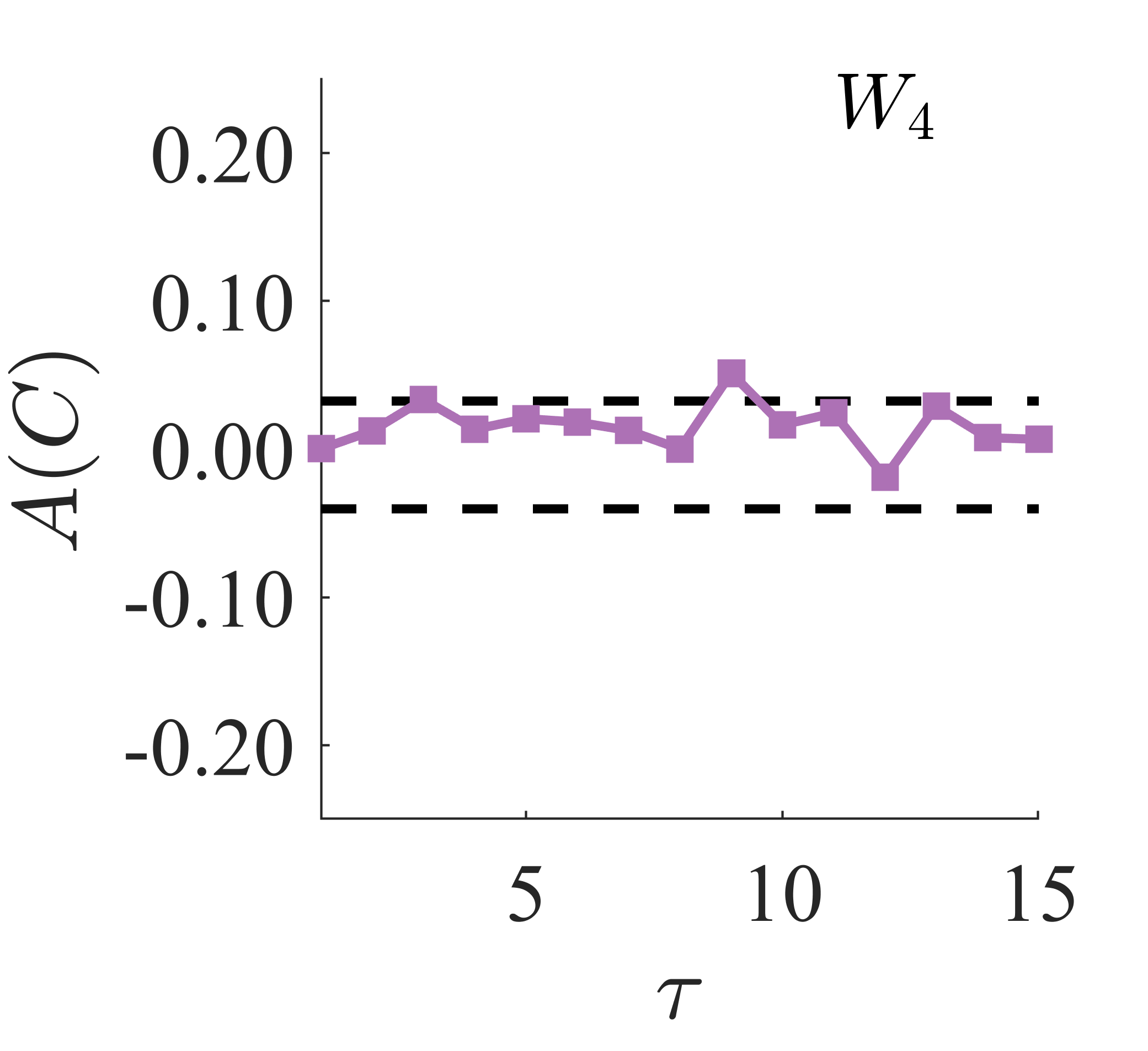}
\end{minipage}
 \begin{minipage}{4cm}
  \begin{flushleft}(b)%
\end{flushleft}%
\includegraphics[width=0.99\columnwidth,clip]{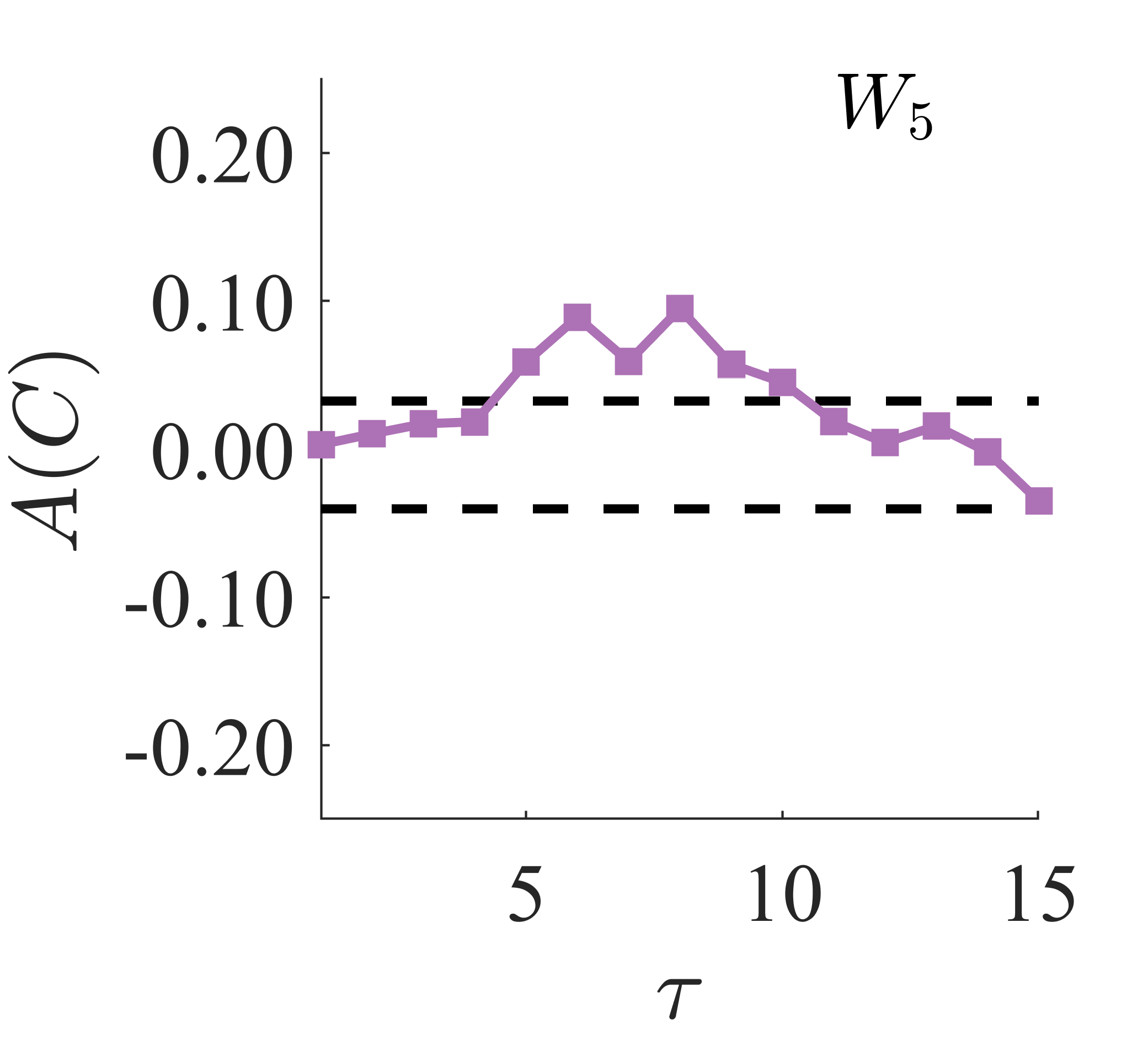}
\end{minipage}
 \begin{minipage}{4cm}
  \begin{flushleft}(c)%
\end{flushleft}%
\includegraphics[width=0.99\columnwidth,clip]{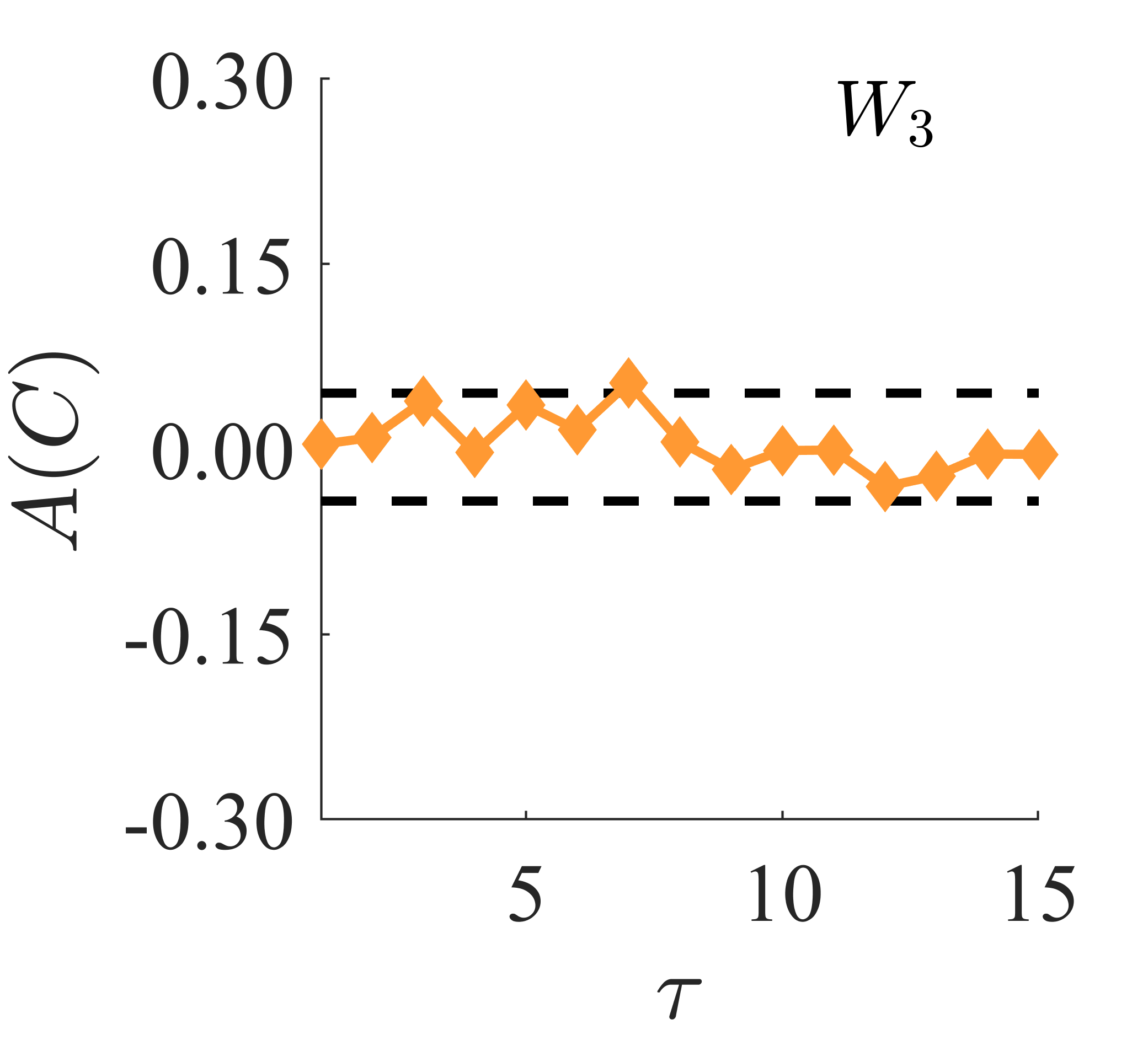}
\end{minipage}
 \begin{minipage}{4cm}
  \begin{flushleft}(d)%
\end{flushleft}%
\includegraphics[width=0.99\columnwidth,clip]{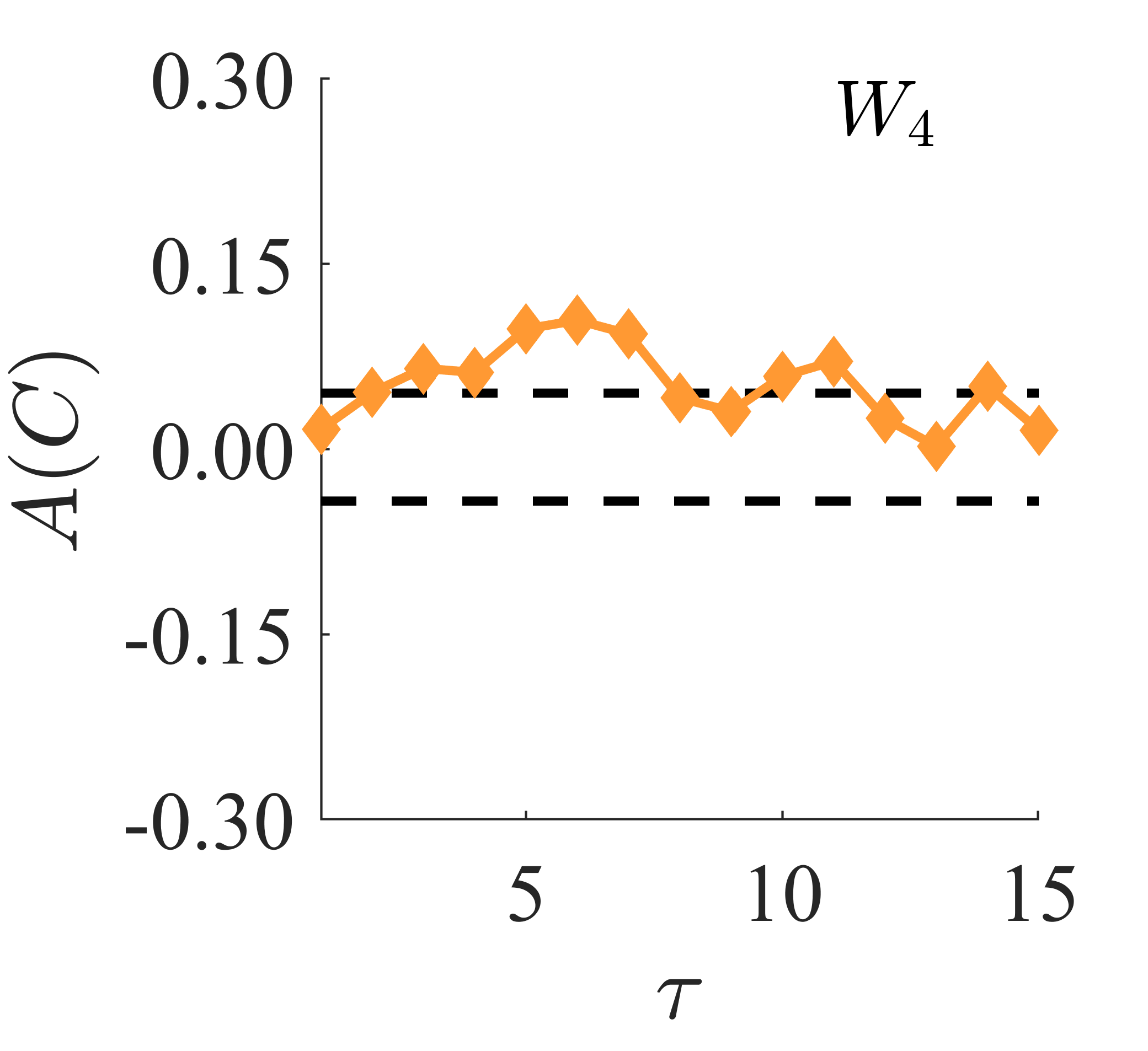}
\end{minipage}
\begin{minipage}{4cm}
\begin{flushleft}(e)%
\end{flushleft}%
\includegraphics[width=0.99\columnwidth,clip]{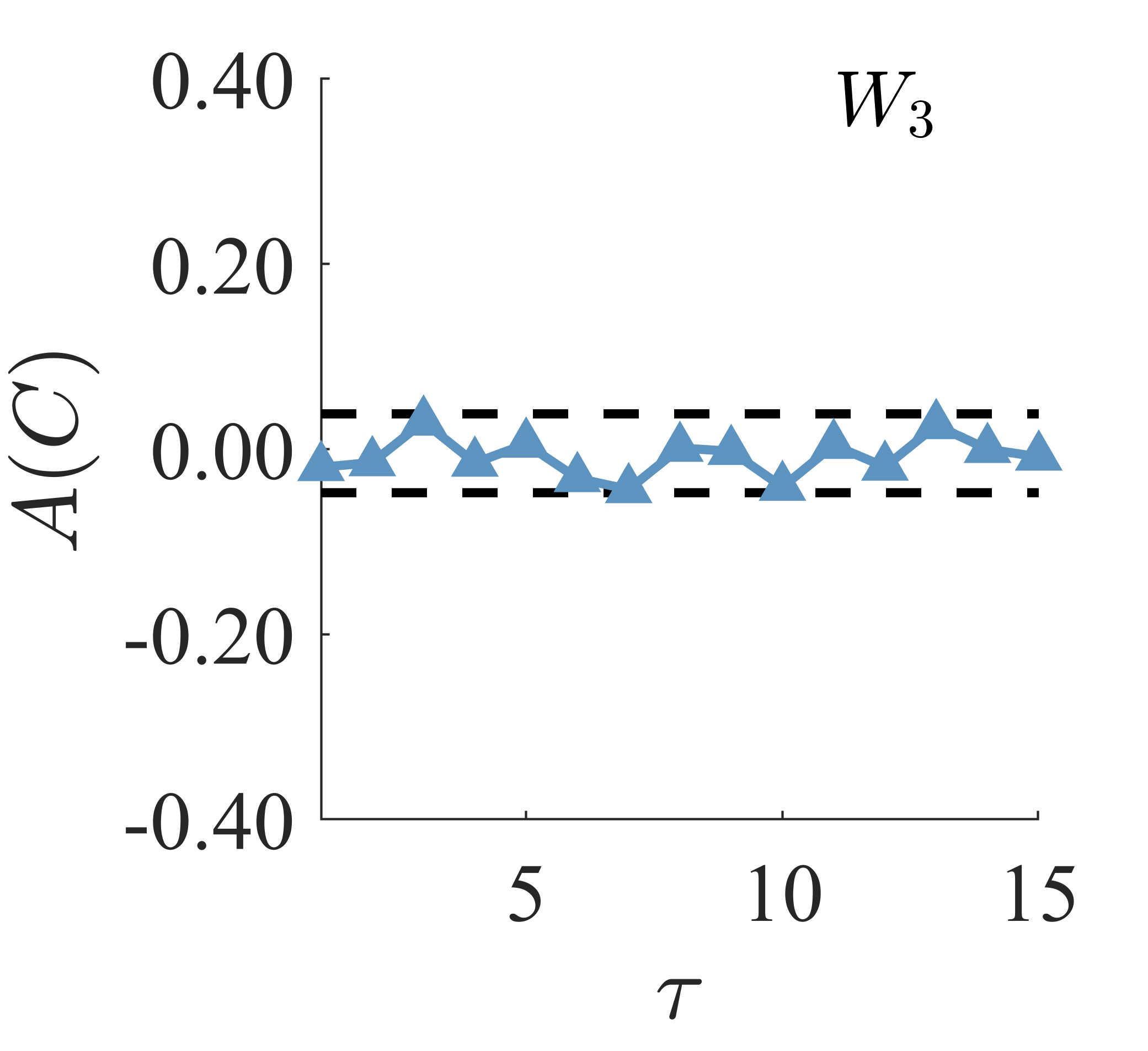}
\end{minipage}
\begin{minipage}{4cm}
\begin{flushleft}(f)%
\end{flushleft}%
\includegraphics[width=0.99\columnwidth,clip]{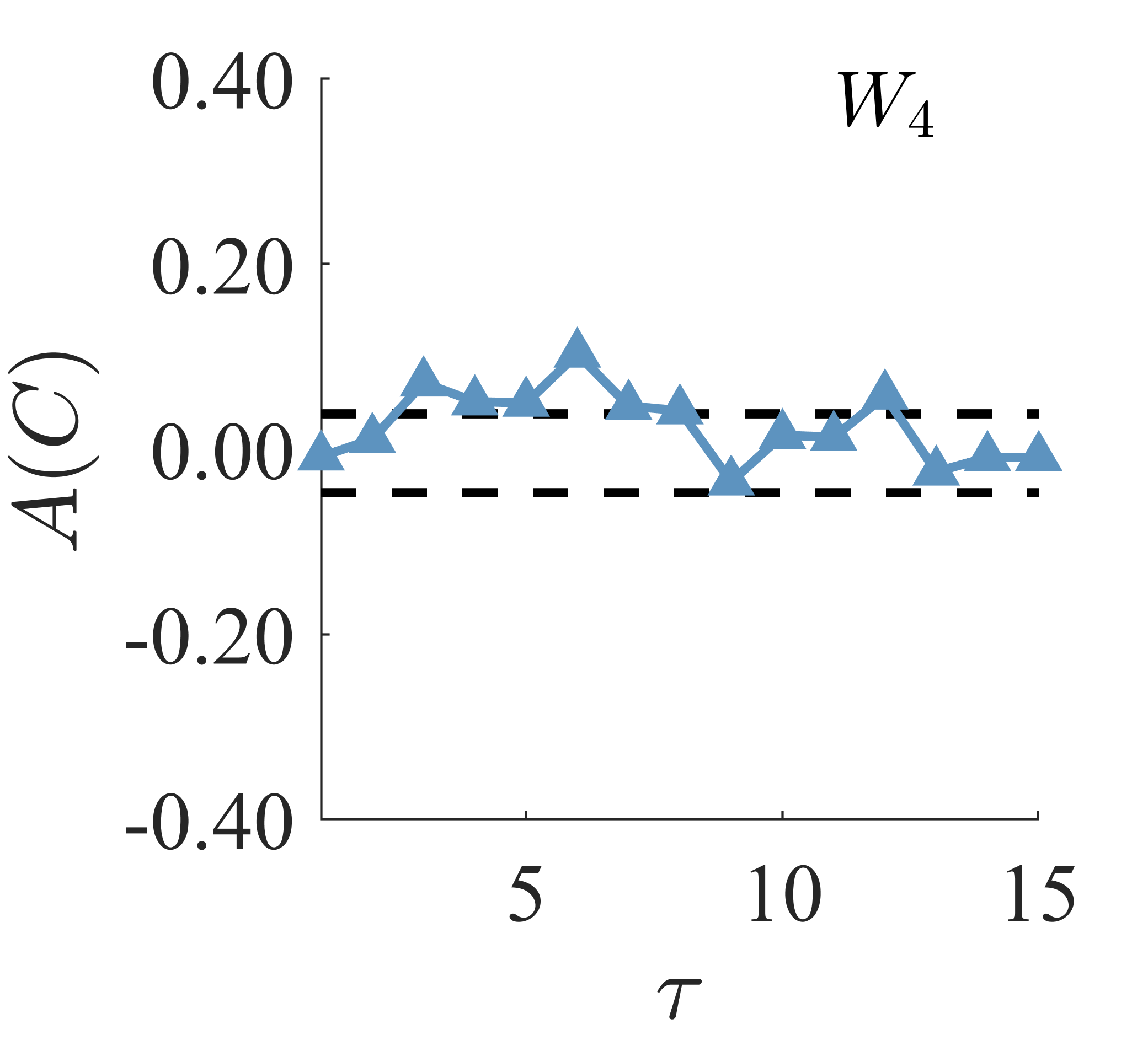}
\end{minipage}
\caption{\label{fig:firstWAC}
Determining the time interval in which the significant separation between Go and No-Go trials can be observed for each region.
Asymmetry index for complexity $A(C)$ as a function of the time delay $\tau$ for the first window showing significant asymmetry index and their previous one: channel 1 (a) $W_4$, (b) $W_5$; channel 3 (c) $W_3$, (d) $W_4$; channel 4 (e) $W_3$, (f) $W_4$.}
\end{figure}%

By using the average event-related potentials time series, Ledberg et al.~\cite{Ledberg07} have found that the significant separation between Go and No-Go trials just started after $t>300$~ms for site 2 which is in agreement with our results since $W_6$ goes up to $340$~ms. Their findings also corroborate the fact that our asymmetry indices are much larger in $W_7$ which comprises the largest interval of significant differences in the average activity.

The advantage of using the multi-scale approach can be verified for example for region 2 during $W_6$ and $W_7$. The largest differences for complexity between Go and No-Go trials does not occur for $\tau=1$ but for $3<\tau<6$ at $W_7$ (and $\tau=3,4,6,9$ at $W_6$).
This strongly suggests that the relevant information for the task is related to a time scale from $15$ to $30$~ms.

In Fig.~\ref{fig:firstWAC} we show the complexity asymmetry index \textit{versus} time delay for $W_n$ and $W_{n-1}$ for sites 1, 3 and 4, where $W_n$ is
the first window showing significant asymmetry index  for each region. 
Regarding these plots, one can see that the parietal areas present specific-response differences already at the interval $150 <t< 240$~ms ($W_4$), whereas the motor cortex just shows significant differences for $t>200$~ms ($W_5$).
It is worth mentioning that Ledberg et al.~\cite{Ledberg07} have reported that the response-specific differences in the average potential start only for
$t>250$~ms at parietal areas 3 and 4. 
This means that by using the information theory quantifiers we can capture significant differences a little earlier in time than with the average activity. 
 
 \begin{center}
 \begin{figure*}[!ht]
 \centering
  \begin{minipage}{8cm}
   \begin{flushleft}(a)
 \end{flushleft}
 \includegraphics[width=0.98\columnwidth,clip]{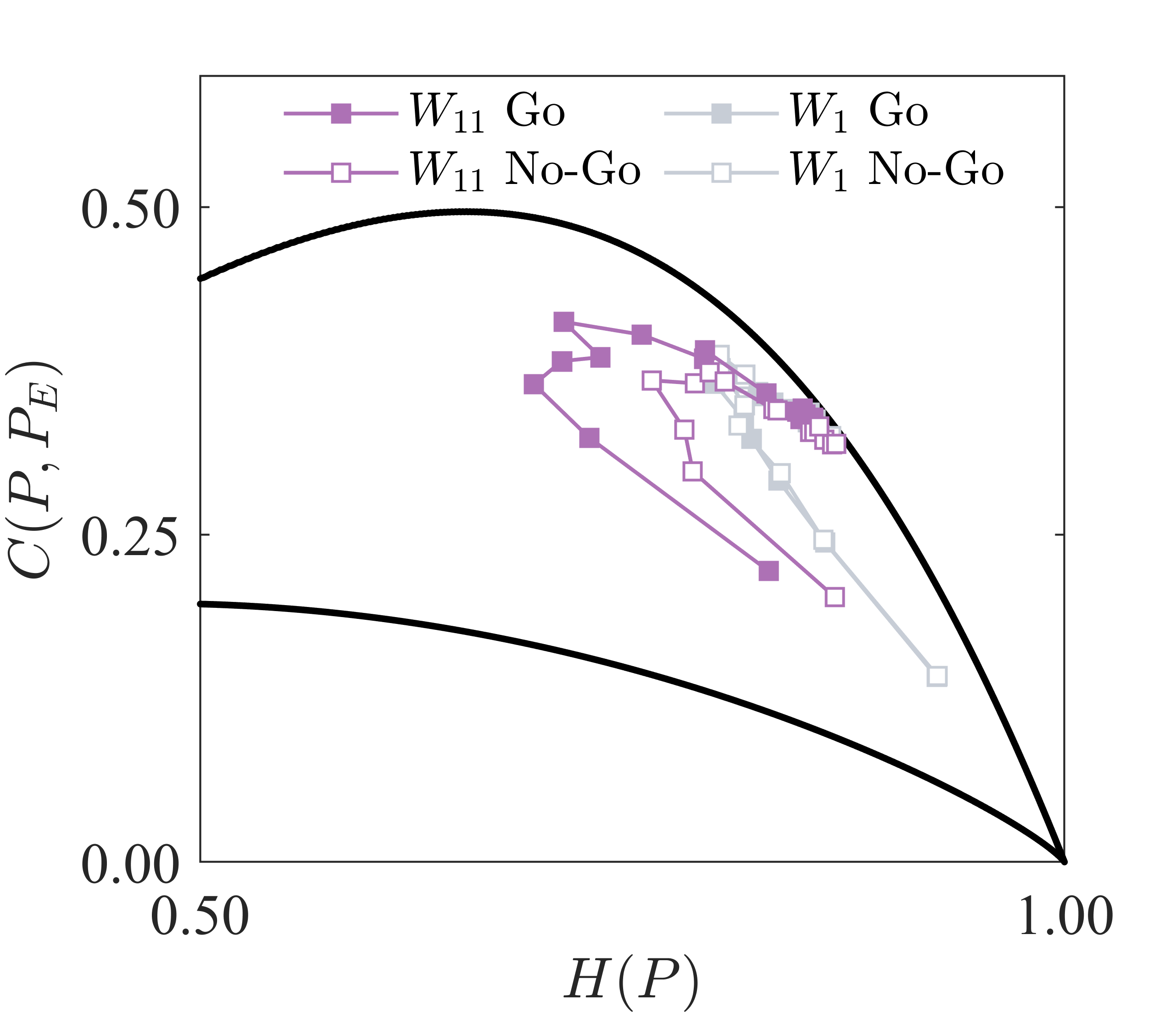}
 \end{minipage}
 \begin{minipage}{8cm}
 \begin{flushleft}(b)
\end{flushleft}
 \includegraphics[width=0.98\columnwidth,clip]{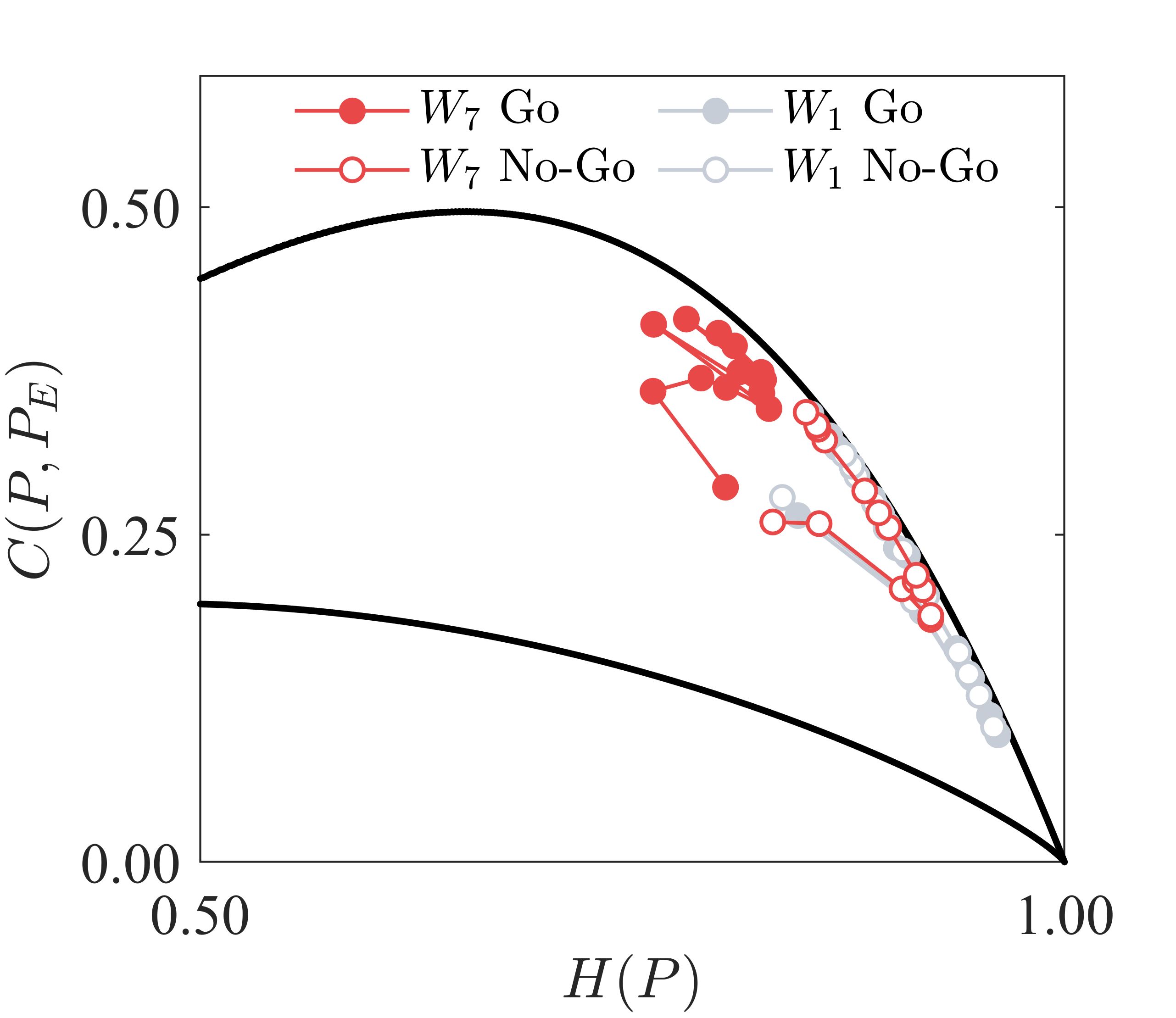}
 \end{minipage}
  \begin{minipage}{8cm}
   \begin{flushleft}(c)
 \end{flushleft}
 \includegraphics[width=0.98\columnwidth,clip]{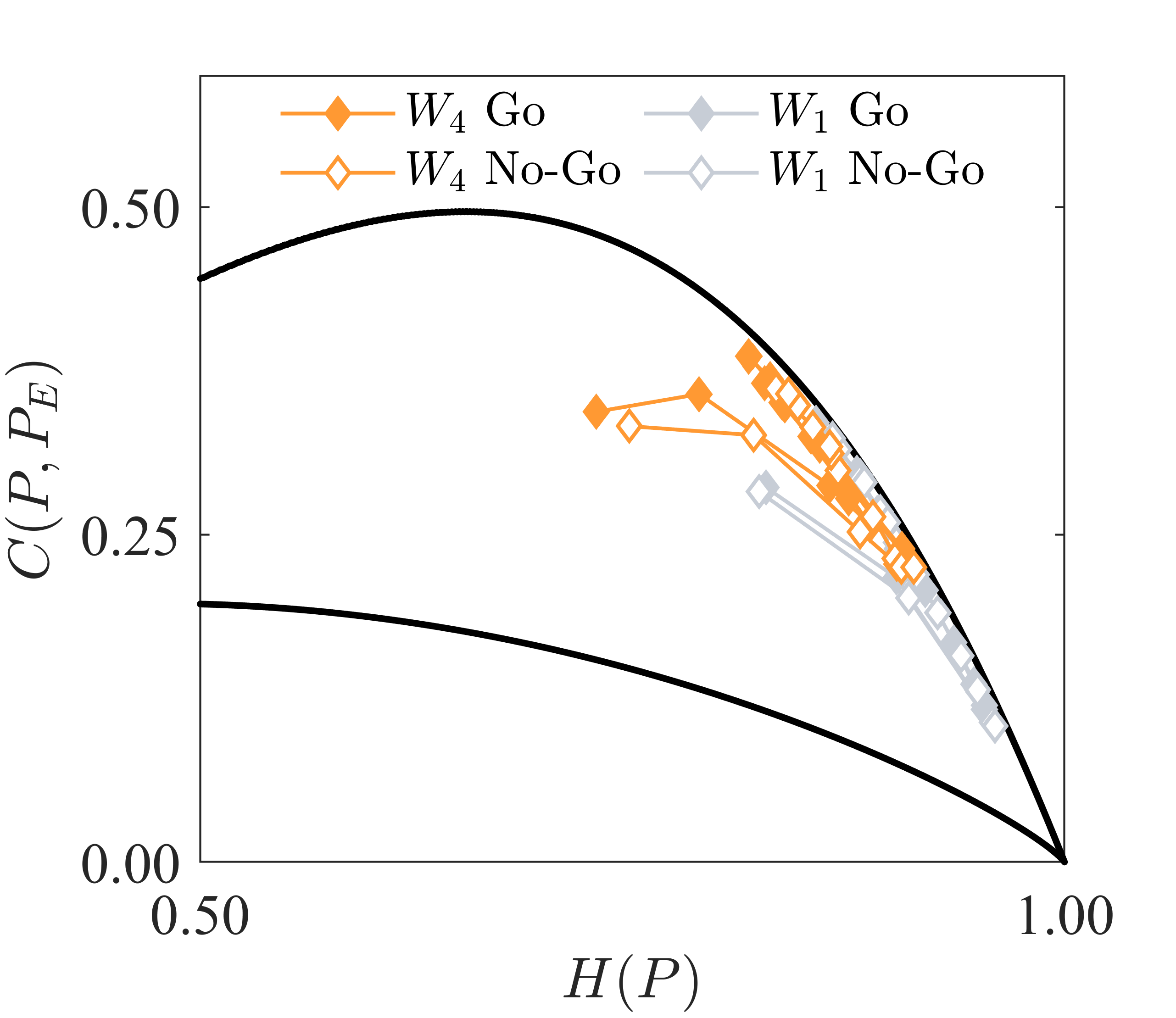}
 \end{minipage}
 \begin{minipage}{8cm}
 \begin{flushleft}(d)
 \end{flushleft}
\includegraphics[width=0.98\columnwidth,clip]{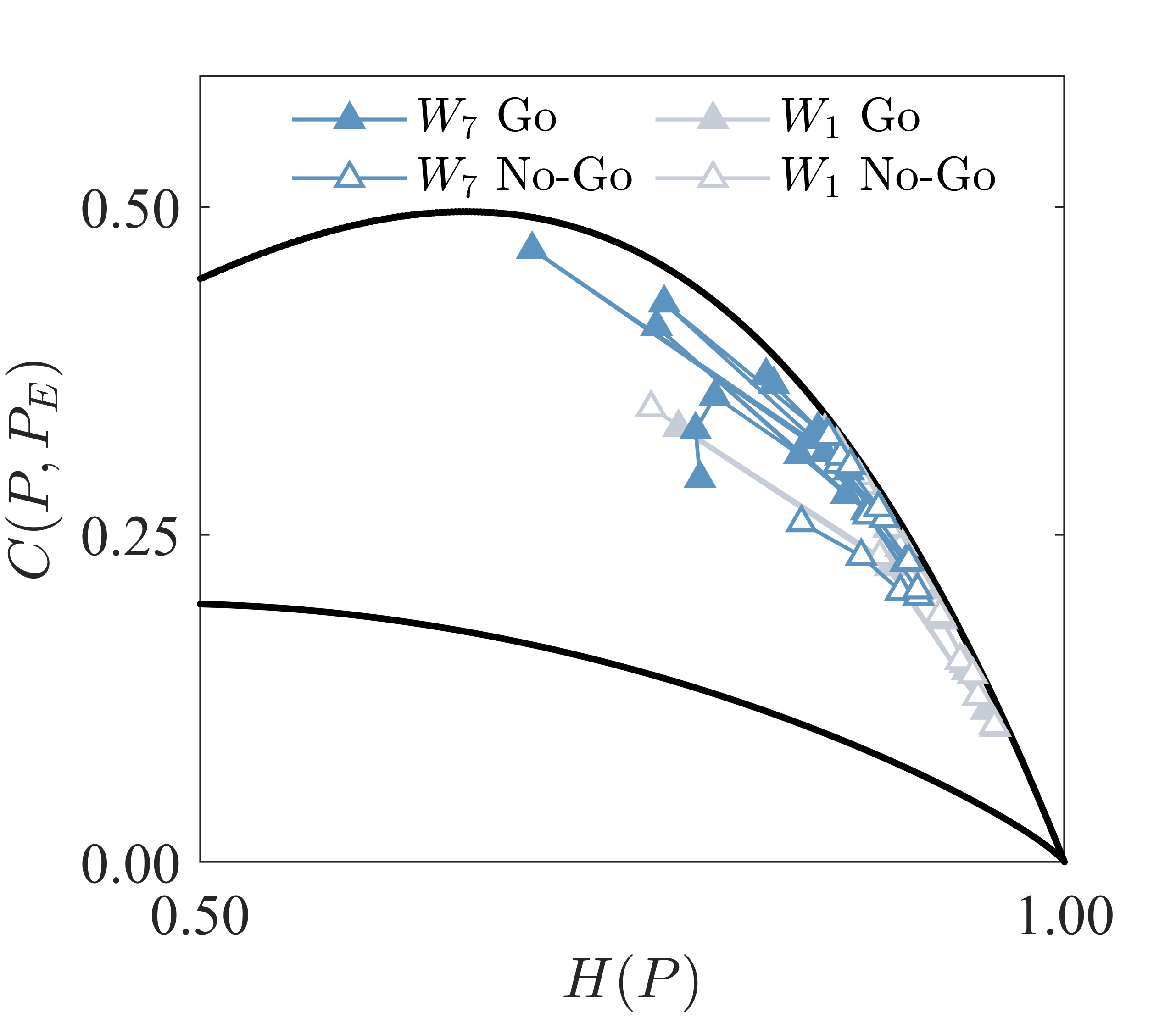}
\end{minipage}
\caption{\label{fig:CxHplane} 
Multi-scale Complexity-Entropy plane for the first 15 time delays $\tau$. For each region we depict $C$ \textit{versus} $H$ values for the first window $W_1$ (Go and No-Go trials in grey) and a post-stimulus window that shows an illustrative separation between Go and No-Go trials in the plane. (a) Region 1: $W_{11}$. (b) Region 2: $W_7$. (c) Region 3: $W_4$. (d) Region 4: $W_7$. Solid lines in black represent the maximum and minimum complexity values for a fixed value of the entropy.}
\end{figure*}
\end{center}

Finally, we can use the multi-scale entropy-complexity plane as an extra way to visualize the separation between Go and No-Go trials at specific time windows (see colored lines in Fig.\ref{fig:CxHplane}). 
Again, the most pronounced difference in the plane happens for region 2 and $W_7$ (Fig.\ref{fig:CxHplane}(b)).
We can also use the $C \times H$ plane to see differences among time windows: by comparing colored lines with grey lines for $W_1$ in the plots, which could mean that the brain is processing information related to the task even if it is not differentiating Go and No-Go trials.
Moreover, we can compare differences among regions: the statistical properties of the motor cortex area is clearly different from the others.
For regions 2, 3 and 4, the system starts close to a totally disordered regime ($H=1$ and $C=0$) and presents an increase in complexity in the direction of $C_{\mathrm{max}}$. The motor area, however, is never close to the border neither to the most disordered state.

 It has been shown by Zhang et al. ~\cite{Zhang08} that for the No-Go trials the oscillatory activity at the beta band reappears at the end of trials. They suggested that since for the No-Go condition the lever pressure was maintained until the end of the recorded time period, the beta rebound reflects the resumption of the same network of the beginning of the trial in support of sensorimotor integration, or preparation for the next trial. In Fig.\ref{fig:CxHplane}(a) we can see for the motor area that despite statistical properties are not the same at $W_1$ and $W_{11}$, the No-Go curves at $W_{11}$ are more similar to $W_1$ than the Go plots, which should be related to the rebound effect.

\section{Conclusion}
\label{Sec:Conclusions}
To summarize, we have shown that information theory quantifiers (such as Shannon entropy, MPR-statistical complexity, and multi-scale entropy-complexity plane) are a useful tool to characterize the information process in the brain signals. On one hand, we can determine cortical regions in which the response-specific information is processed. On the other hand, we can estimate time intervals and time scales in which these differences are more pronounced. 

By introducing the asymmetry index we can statically quantify the differences between Go and No-Go trials.
First, by combining entropy and complexity we can determine in which interval of the trial it is possible to verify any difference between trial types. For some intervals, only one of the two measures presents a significant difference between Go and No-Go trials which means that it is important to calculate both.
Second, by using different values of the time embedding we can find response-specific differences earlier in time than by using the average potential. We can also estimate what are the important time scales of the information process by verifying which time delays maximizes the asymmetry index.

This means that the information theory quantifiers and their asymmetry index
can be employed together with average event-related potentials to characterize response-specific brain activity. Our results open new avenues in the investigation of response-specific or stimulus-specific brain activity. Moreover, the method is potentially useful to quantify other features of the task such as differences between intervals of time and between regions.


\begin{acknowledgements}
The authors thank FAPEAL, UFAL, CNPq (grant 432429/2016-6) and CAPES 
(grant 88881.120309/2016-01) for financial support.
\end{acknowledgements}
\bibliography{matias}

\begin{thebibliography}{23}
\expandafter\ifx\csname natexlab\endcsname\relax\def\natexlab#1{#1}\fi
\expandafter\ifx\csname bibnamefont\endcsname\relax
  \def\bibnamefont#1{#1}\fi
\expandafter\ifx\csname bibfnamefont\endcsname\relax
  \def\bibfnamefont#1{#1}\fi
\expandafter\ifx\csname citenamefont\endcsname\relax
  \def\citenamefont#1{#1}\fi
\expandafter\ifx\csname url\endcsname\relax
  \def\url#1{\texttt{#1}}\fi
\expandafter\ifx\csname urlprefix\endcsname\relax\def\urlprefix{URL }\fi
\providecommand{\bibinfo}[2]{#2}
\providecommand{\eprint}[2][]{\url{#2}}

\bibitem[{\citenamefont{Sporns}(2011)}]{Sporns11}
\bibinfo{editor}{\bibfnamefont{O.}~\bibnamefont{Sporns}}, ed.,
  \emph{\bibinfo{title}{Networks of the brain}} (\bibinfo{publisher}{The MIT
  Press (Cambridge)}, \bibinfo{year}{2011}).

\bibitem[{\citenamefont{Bressler et~al.}(1993)\citenamefont{Bressler, Coppola,
  and Nakamura}}]{Bressler93}
\bibinfo{author}{\bibfnamefont{S.~L.} \bibnamefont{Bressler}},
  \bibinfo{author}{\bibfnamefont{R.}~\bibnamefont{Coppola}}, \bibnamefont{and}
  \bibinfo{author}{\bibfnamefont{R.}~\bibnamefont{Nakamura}},
  \bibinfo{journal}{Nature} \textbf{\bibinfo{volume}{366}},
  \bibinfo{pages}{153} (\bibinfo{year}{1993}).

\bibitem[{\citenamefont{Liang et~al.}(2000)\citenamefont{Liang, Ding, Nakamura,
  and Bressler}}]{Liang00}
\bibinfo{author}{\bibfnamefont{H.}~\bibnamefont{Liang}},
  \bibinfo{author}{\bibfnamefont{M.}~\bibnamefont{Ding}},
  \bibinfo{author}{\bibfnamefont{R.}~\bibnamefont{Nakamura}}, \bibnamefont{and}
  \bibinfo{author}{\bibfnamefont{S.~L.} \bibnamefont{Bressler}},
  \bibinfo{journal}{Neuroreport} \textbf{\bibinfo{volume}{11}},
  \bibinfo{pages}{2875} (\bibinfo{year}{2000}).

\bibitem[{\citenamefont{Brovelli et~al.}(2004)\citenamefont{Brovelli, Ding,
  Ledberg, Chen, Nakamura, and Bressler}}]{Brovelli04}
\bibinfo{author}{\bibfnamefont{A.}~\bibnamefont{Brovelli}},
  \bibinfo{author}{\bibfnamefont{M.}~\bibnamefont{Ding}},
  \bibinfo{author}{\bibfnamefont{A.}~\bibnamefont{Ledberg}},
  \bibinfo{author}{\bibfnamefont{Y.}~\bibnamefont{Chen}},
  \bibinfo{author}{\bibfnamefont{R.}~\bibnamefont{Nakamura}}, \bibnamefont{and}
  \bibinfo{author}{\bibfnamefont{S.~L.} \bibnamefont{Bressler}},
  \bibinfo{journal}{Proc. Natl. Acad. Sci. {USA}}
  \textbf{\bibinfo{volume}{101}}, \bibinfo{pages}{9849} (\bibinfo{year}{2004}).

\bibitem[{\citenamefont{Salazar et~al.}(2012)\citenamefont{Salazar, Dotson,
  Bressler, and Gray}}]{Salazar12}
\bibinfo{author}{\bibfnamefont{R.~F.} \bibnamefont{Salazar}},
  \bibinfo{author}{\bibfnamefont{N.~M.} \bibnamefont{Dotson}},
  \bibinfo{author}{\bibfnamefont{S.~L.} \bibnamefont{Bressler}},
  \bibnamefont{and} \bibinfo{author}{\bibfnamefont{C.~M.} \bibnamefont{Gray}},
  \bibinfo{journal}{Science} \textbf{\bibinfo{volume}{338}},
  \bibinfo{pages}{1097} (\bibinfo{year}{2012}).

\bibitem[{\citenamefont{Dotson et~al.}(2014)\citenamefont{Dotson, Salazar, and
  Gray}}]{Dotson14}
\bibinfo{author}{\bibfnamefont{N.~M.} \bibnamefont{Dotson}},
  \bibinfo{author}{\bibfnamefont{R.~F.} \bibnamefont{Salazar}},
  \bibnamefont{and} \bibinfo{author}{\bibfnamefont{C.~M.} \bibnamefont{Gray}},
  \bibinfo{journal}{The Journal of Neuroscience} \textbf{\bibinfo{volume}{34}},
  \bibinfo{pages}{13600} (\bibinfo{year}{2014}).

\bibitem[{\citenamefont{Ledberg et~al.}(2007)\citenamefont{Ledberg, Bressler,
  Ding, Coppola, and Nakamura}}]{Ledberg07}
\bibinfo{author}{\bibfnamefont{A.}~\bibnamefont{Ledberg}},
  \bibinfo{author}{\bibfnamefont{S.~L.} \bibnamefont{Bressler}},
  \bibinfo{author}{\bibfnamefont{M.}~\bibnamefont{Ding}},
  \bibinfo{author}{\bibfnamefont{R.}~\bibnamefont{Coppola}}, \bibnamefont{and}
  \bibinfo{author}{\bibfnamefont{R.}~\bibnamefont{Nakamura}},
  \bibinfo{journal}{Cerebral cortex} \textbf{\bibinfo{volume}{17}},
  \bibinfo{pages}{44} (\bibinfo{year}{2007}).

\bibitem[{\citenamefont{Shannon and Weaver}(1949)}]{Shannon49}
\bibinfo{author}{\bibfnamefont{C.}~\bibnamefont{Shannon}} \bibnamefont{and}
  \bibinfo{author}{\bibfnamefont{W.}~\bibnamefont{Weaver}},
  \emph{\bibinfo{title}{The mathematical theory of communication}}
  (\bibinfo{publisher}{Champaign, IL: University of Illinois Press},
  \bibinfo{year}{1949}).

\bibitem[{\citenamefont{Lamberti et~al.}(2004)\citenamefont{Lamberti,
  Mart\'{\i}n, Plastino, and Rosso}}]{Lamberti04}
\bibinfo{author}{\bibfnamefont{P.~W.} \bibnamefont{Lamberti}},
  \bibinfo{author}{\bibfnamefont{M.~T.} \bibnamefont{Mart\'{\i}n}},
  \bibinfo{author}{\bibfnamefont{A.}~\bibnamefont{Plastino}}, \bibnamefont{and}
  \bibinfo{author}{\bibfnamefont{O.~A.} \bibnamefont{Rosso}},
  \bibinfo{journal}{Physica A: Statistical Mechanics and its Applications}
  \textbf{\bibinfo{volume}{334}}, \bibinfo{pages}{119} (\bibinfo{year}{2004}).

\bibitem[{\citenamefont{Rosso et~al.}(2007)\citenamefont{Rosso, Larrondo,
  Mart\'{\i}n, Plastino, and Fuentes}}]{Rosso07}
\bibinfo{author}{\bibfnamefont{O.~A.} \bibnamefont{Rosso}},
  \bibinfo{author}{\bibfnamefont{H.~A.} \bibnamefont{Larrondo}},
  \bibinfo{author}{\bibfnamefont{M.~T.} \bibnamefont{Mart\'{\i}n}},
  \bibinfo{author}{\bibfnamefont{A.}~\bibnamefont{Plastino}}, \bibnamefont{and}
  \bibinfo{author}{\bibfnamefont{M.}~\bibnamefont{Fuentes}},
  \bibinfo{journal}{Physical Review Letters} \textbf{\bibinfo{volume}{99}},
  \bibinfo{pages}{154102} (\bibinfo{year}{2007}).

\bibitem[{\citenamefont{Zunino et~al.}(2012)\citenamefont{Zunino, Soriano, and
  Rosso}}]{Zunino12}
\bibinfo{author}{\bibfnamefont{L.}~\bibnamefont{Zunino}},
  \bibinfo{author}{\bibfnamefont{M.~C.} \bibnamefont{Soriano}},
  \bibnamefont{and} \bibinfo{author}{\bibfnamefont{O.~A.} \bibnamefont{Rosso}},
  \bibinfo{journal}{Phys. Rev. E} \textbf{\bibinfo{volume}{86}},
  \bibinfo{pages}{046210} (\bibinfo{year}{2012}).

\bibitem[{\citenamefont{Xiong et~al.}(2020)\citenamefont{Xiong, Shang, He, and
  Zhang}}]{Chinitos20}
\bibinfo{author}{\bibfnamefont{H.}~\bibnamefont{Xiong}},
  \bibinfo{author}{\bibfnamefont{P.}~\bibnamefont{Shang}},
  \bibinfo{author}{\bibfnamefont{J.}~\bibnamefont{He}}, \bibnamefont{and}
  \bibinfo{author}{\bibfnamefont{Y.}~\bibnamefont{Zhang}},
  \bibinfo{journal}{Nonlinear Dynamics} \textbf{\bibinfo{volume}{100}},
  \bibinfo{pages}{1673–1687} (\bibinfo{year}{2020}).

\bibitem[{\citenamefont{Bandt and Pompe}(2002)}]{Bandt02}
\bibinfo{author}{\bibfnamefont{C.}~\bibnamefont{Bandt}} \bibnamefont{and}
  \bibinfo{author}{\bibfnamefont{B.}~\bibnamefont{Pompe}},
  \bibinfo{journal}{Physical review letters} \textbf{\bibinfo{volume}{88}},
  \bibinfo{pages}{174102} (\bibinfo{year}{2002}).

\bibitem[{\citenamefont{Montani
  et~al.}(2015{\natexlab{a}})\citenamefont{Montani, Rosso, Matias, Bressler,
  and Mirasso}}]{Montani15}
\bibinfo{author}{\bibfnamefont{F.}~\bibnamefont{Montani}},
  \bibinfo{author}{\bibfnamefont{O.~A.} \bibnamefont{Rosso}},
  \bibinfo{author}{\bibfnamefont{F.~S.} \bibnamefont{Matias}},
  \bibinfo{author}{\bibfnamefont{S.~L.} \bibnamefont{Bressler}},
  \bibnamefont{and} \bibinfo{author}{\bibfnamefont{C.~R.}
  \bibnamefont{Mirasso}}, \bibinfo{journal}{Phil. Trans. R. Soc. A}
  \textbf{\bibinfo{volume}{373}}, \bibinfo{pages}{20150110}
  (\bibinfo{year}{2015}{\natexlab{a}}).

\bibitem[{\citenamefont{Lotfi et~al.}(2020)\citenamefont{Lotfi, Feliciano,
  Aguiar, Silva, Carvalho, Rosso, Copelli, Matias, and Carelli}}]{Lotfi2020b}
\bibinfo{author}{\bibfnamefont{N.}~\bibnamefont{Lotfi}},
  \bibinfo{author}{\bibfnamefont{T.}~\bibnamefont{Feliciano}},
  \bibinfo{author}{\bibfnamefont{L.~A.} \bibnamefont{Aguiar}},
  \bibinfo{author}{\bibfnamefont{T.~P.~L.} \bibnamefont{Silva}},
  \bibinfo{author}{\bibfnamefont{T.~T.} \bibnamefont{Carvalho}},
  \bibinfo{author}{\bibfnamefont{O.~A.} \bibnamefont{Rosso}},
  \bibinfo{author}{\bibfnamefont{M.}~\bibnamefont{Copelli}},
  \bibinfo{author}{\bibfnamefont{F.~S.} \bibnamefont{Matias}},
  \bibnamefont{and} \bibinfo{author}{\bibfnamefont{P.~V.}
  \bibnamefont{Carelli}}, \bibinfo{journal}{arXiv preprint arXiv:2010.04123}
  (\bibinfo{year}{2020}).

\bibitem[{\citenamefont{Rosso et~al.}(2006)\citenamefont{Rosso, Martin,
  Figliola, Keller, and Plastino}}]{Rosso06}
\bibinfo{author}{\bibfnamefont{O.}~\bibnamefont{Rosso}},
  \bibinfo{author}{\bibfnamefont{M.}~\bibnamefont{Martin}},
  \bibinfo{author}{\bibfnamefont{A.}~\bibnamefont{Figliola}},
  \bibinfo{author}{\bibfnamefont{K.}~\bibnamefont{Keller}}, \bibnamefont{and}
  \bibinfo{author}{\bibfnamefont{A.}~\bibnamefont{Plastino}},
  \bibinfo{journal}{Journal of neuroscience methods}
  \textbf{\bibinfo{volume}{153}}, \bibinfo{pages}{163} (\bibinfo{year}{2006}).

\bibitem[{\citenamefont{Montani
  et~al.}(2015{\natexlab{b}})\citenamefont{Montani, Baravalle, Montangie, and
  Rosso}}]{Montani15b}
\bibinfo{author}{\bibfnamefont{F.}~\bibnamefont{Montani}},
  \bibinfo{author}{\bibfnamefont{R.}~\bibnamefont{Baravalle}},
  \bibinfo{author}{\bibfnamefont{L.}~\bibnamefont{Montangie}},
  \bibnamefont{and} \bibinfo{author}{\bibfnamefont{O.~A.} \bibnamefont{Rosso}},
  \bibinfo{journal}{Philosophical Transactions of the Royal Society A:
  Mathematical, Physical and Engineering Sciences}
  \textbf{\bibinfo{volume}{373}}, \bibinfo{pages}{20150109}
  (\bibinfo{year}{2015}{\natexlab{b}}).

\bibitem[{\citenamefont{Montani et~al.}(2014)\citenamefont{Montani, Deleglise,
  and Rosso}}]{Montani14}
\bibinfo{author}{\bibfnamefont{F.}~\bibnamefont{Montani}},
  \bibinfo{author}{\bibfnamefont{E.~B.} \bibnamefont{Deleglise}},
  \bibnamefont{and} \bibinfo{author}{\bibfnamefont{O.~A.} \bibnamefont{Rosso}},
  \bibinfo{journal}{Physica A: Statistical Mechanics and its Applications}
  \textbf{\bibinfo{volume}{401}}, \bibinfo{pages}{58} (\bibinfo{year}{2014}).

\bibitem[{\citenamefont{R.~López-Ruiz and Calbet}(1995)}]{LMC95}
\bibinfo{author}{\bibfnamefont{H.~M.} \bibnamefont{R.~López-Ruiz}}
  \bibnamefont{and} \bibinfo{author}{\bibfnamefont{X.}~\bibnamefont{Calbet}},
  \bibinfo{journal}{Physics Letters A} \textbf{\bibinfo{volume}{209}},
  \bibinfo{pages}{321 } (\bibinfo{year}{1995}).

\bibitem[{\citenamefont{Grosse et~al.}(2002)\citenamefont{Grosse,
  Bernaola-Galv\'an, Carpena, Rom\'an-Rold\'an, Oliver, and
  Stanley}}]{Grosse02}
\bibinfo{author}{\bibfnamefont{I.}~\bibnamefont{Grosse}},
  \bibinfo{author}{\bibfnamefont{P.}~\bibnamefont{Bernaola-Galv\'an}},
  \bibinfo{author}{\bibfnamefont{P.}~\bibnamefont{Carpena}},
  \bibinfo{author}{\bibfnamefont{R.}~\bibnamefont{Rom\'an-Rold\'an}},
  \bibinfo{author}{\bibfnamefont{J.}~\bibnamefont{Oliver}}, \bibnamefont{and}
  \bibinfo{author}{\bibfnamefont{H.~E.} \bibnamefont{Stanley}},
  \bibinfo{journal}{Phys. Rev. E} \textbf{\bibinfo{volume}{65}},
  \bibinfo{pages}{041905} (\bibinfo{year}{2002}).

\bibitem[{\citenamefont{M.T.~Martin and Rosso}(2006)}]{MPR-cotas}
\bibinfo{author}{\bibfnamefont{A.~P.} \bibnamefont{M.T.~Martin}}
  \bibnamefont{and} \bibinfo{author}{\bibfnamefont{O.}~\bibnamefont{Rosso}},
  \bibinfo{journal}{Physica A: Statistical Mechanics and its Applications}
  \textbf{\bibinfo{volume}{369}}, \bibinfo{pages}{439 } (\bibinfo{year}{2006}).

\bibitem[{\citenamefont{Matias et~al.}(2014)\citenamefont{Matias, Gollo,
  Carelli, Bressler, Copelli, and Mirasso}}]{Matias14}
\bibinfo{author}{\bibfnamefont{F.~S.} \bibnamefont{Matias}},
  \bibinfo{author}{\bibfnamefont{L.~L.} \bibnamefont{Gollo}},
  \bibinfo{author}{\bibfnamefont{P.~V.} \bibnamefont{Carelli}},
  \bibinfo{author}{\bibfnamefont{S.~L.} \bibnamefont{Bressler}},
  \bibinfo{author}{\bibfnamefont{M.}~\bibnamefont{Copelli}}, \bibnamefont{and}
  \bibinfo{author}{\bibfnamefont{C.~R.} \bibnamefont{Mirasso}},
  \bibinfo{journal}{NeuroImage} \textbf{\bibinfo{volume}{99}},
  \bibinfo{pages}{411} (\bibinfo{year}{2014}).

\bibitem[{\citenamefont{Zhang et~al.}(2008)\citenamefont{Zhang, Chen, Bressler,
  and Ding}}]{Zhang08}
\bibinfo{author}{\bibfnamefont{Y.}~\bibnamefont{Zhang}},
  \bibinfo{author}{\bibfnamefont{Y.}~\bibnamefont{Chen}},
  \bibinfo{author}{\bibfnamefont{S.~L.} \bibnamefont{Bressler}},
  \bibnamefont{and} \bibinfo{author}{\bibfnamefont{M.}~\bibnamefont{Ding}},
  \bibinfo{journal}{Neuroscience} \textbf{\bibinfo{volume}{156}}
  (\bibinfo{year}{2008}).

\end{thebibliography}
 \end{document}